\documentclass[aps,prl,reprint,amsmath,longbibliography]{revtex4-2}

\usepackage{graphicx}
\usepackage{epstopdf}
\usepackage{amsmath,bm}
\usepackage{threeparttable}

\usepackage{array}
\usepackage{makecell}
\usepackage{booktabs}
\usepackage{longtable}
\usepackage{multirow}
\usepackage{chngcntr}
\usepackage{appendix}
\usepackage{amsmath}

\usepackage{mathrsfs}

\usepackage{hyperref}
\hypersetup{
    colorlinks=true,
    citecolor=blue,
    linkcolor=blue,
}

\usepackage{xcite}  
\usepackage{tabularx}

\newcommand{\Tstrut}{\rule{0pt}{2.6ex}}
\newcommand{\Bstrut}{\rule[-0.9ex]{0pt}{0pt}}
\newcommand{\polbox}[1]{%
	\ensuremath{\left(\!\begin{array}{c}#1\end{array}\!\right)}%
}
\newcommand{\catcelltwo}[1]{\multirow{2}{*}[4mm]{#1}}
\newcommand{\catcellthree}[1]{\multirow{3}{*}[8mm]{#1}}

\begin{document}

\title{Altermagnetic type-II Multiferroics with N\'eel-order-locked Electric Polarization}
	\author{Wen-Ti Guo$^{1}$}
\thanks{These authors contributed equally.}

\author{Junqi Xu$^{1}$}
\thanks{These authors contributed equally.}

\author{Yurong Yang$^{2}$}

\author{Haijun Zhang$^{1,3,4}$}%
\email{zhanghj@nju.edu.cn}
\author{Huaiqiang Wang$^{5,3,4}$}
\email{hqwang@njnu.edu.cn}

\affiliation{1 National Laboratory of Solid State Microstructures, School of Physics, Nanjing University, Nanjing 210093, China
}
\affiliation{2 Jiangsu Key Laboratory of Artificial Functional Materials and Department of Materials Science and Engineering,Nanjing University, Nanjing 210093, China
}
\affiliation{3 Collaborative Innovation Center of Advanced Microstructures, Nanjing University, Nanjing 210093, China
}
\affiliation{4 Jiangsu Physical Science Research Center, Nanjing 210093, China
}
\affiliation{5 Center for Quantum Transport and Thermal Energy Science, School of Physics and Technology, Nanjing Normal University, Nanjing 210023, China
}

\begin{abstract}
	Altermagnetism, an emergent magnetic phase featuring compensated collinear magnetic moments and momentum-dependent spin splittings, has recently garnered widespread interest. A critical issue concerns whether the unconventional spin structures can generate spontaneous electric polarization in altermagnets, thereby achieving type-II multiferroicity. Here, with the combination of symmetry analysis and metal-ligand model, we explicitly demonstrate the generation of electric polarization by altermagnetic N\'eel order.~We further uncover the locking behaviors between N\'eel order and electric polarization, which are classified into eight distinct categories for two-dimensional altermagnets governed by layer group symmetries. Then we take monolayer MgFe$_2$N$_2$ as a prototypical example of altermagnetic type-II multiferroics by first-principles calculations. We also propose to identify the N\'eel order and accompanying electric polarization in altermagnetic multiferroics by magneto-optical microscopy. Bridging type-II multiferroics and altermagnets, our work could pave the way for altermagnetic multifunctional spintronics.
\end{abstract}
\maketitle

\noindent\textbf{Introduction}\\
Altermagnetism, an emerging magnetic phase characterized by compensated collinear magnetic moments and momentum-dependent spin splittings, represents a new class of magnetic materials beyond traditional ferromagnets and antiferromagnets~\cite{ma2021aMultifunctiona, mazin2021prediction, gonzalez2021efficient,smejkal2022emerging, smejkal2022conventional, turek2022altermagnetism, smejkal2022anomalous,  liu2022spin-group, krempasky2024altermagnetic, reimers2024direct, tamang2024newly, liu2024absence, wang2022magneto, ding2024large, chen2024altermagnetic,  zhou2024crystal1, attias2024intrinsic, han2024observation, zeng2024description,  xiao2024spin, jiang2024enumeration, xiao2024anomalous, leeb2024spontaneous, zhang2025Crystal, qian2025fragile, liu2025different, liu2025d-wave, wang2025giant, gao2025ai, yuan2025unconventional, hu2025Catalog, hayami2020bottom}. Uniquely bridging ferromagnetic and antiferromagnetic characteristics, altermagnets could exhibit time-reversal-breaking magneto responses and simultaneously display negligible stray fields and high-frequency spin dynamics, making them promising candidates in ultrafast spintronic devices~\cite{tamang2024newly, bai2024altermagnetism, song2025altermagnets, fender2025altermagnetism, peng2025all}. While recent experimental advances employing techniques such as angle-resolved photoemission spectroscopy~\cite{ding2024large, reimers2024direct, liu2024absence,santhosh2025altermagnetic}, magneto-optical microscopy~\cite{samanta2020crystal, zhang2024probing, raboni2025nanoscale}, and anomalous Hall measurements~\cite{leiviska2024anisotropy, zhou2025manipulation}, have validated the unconventional electronic and magnetic signatures in altermagnets~\cite{wu2025optical,  zhou2025manipulation, han2025discovery}, magnetoelectric (ME) coupling effects of altermagnets remain elusive~\cite{guo2023altermagnetic, zhang2024predictable, matsuda2024multiferroic, sheng2024ubiquitous, gu2025ferroelectric, duan2025antiferroelectric, camerano2025multiferroic, wang2025two}, and critically, the interplay between unconventional magnetic order and emergent electric polarization remains largely unexplored.

\begin{figure}[!htbp]
	\begin{centering}
		\includegraphics[width=0.45\textwidth]{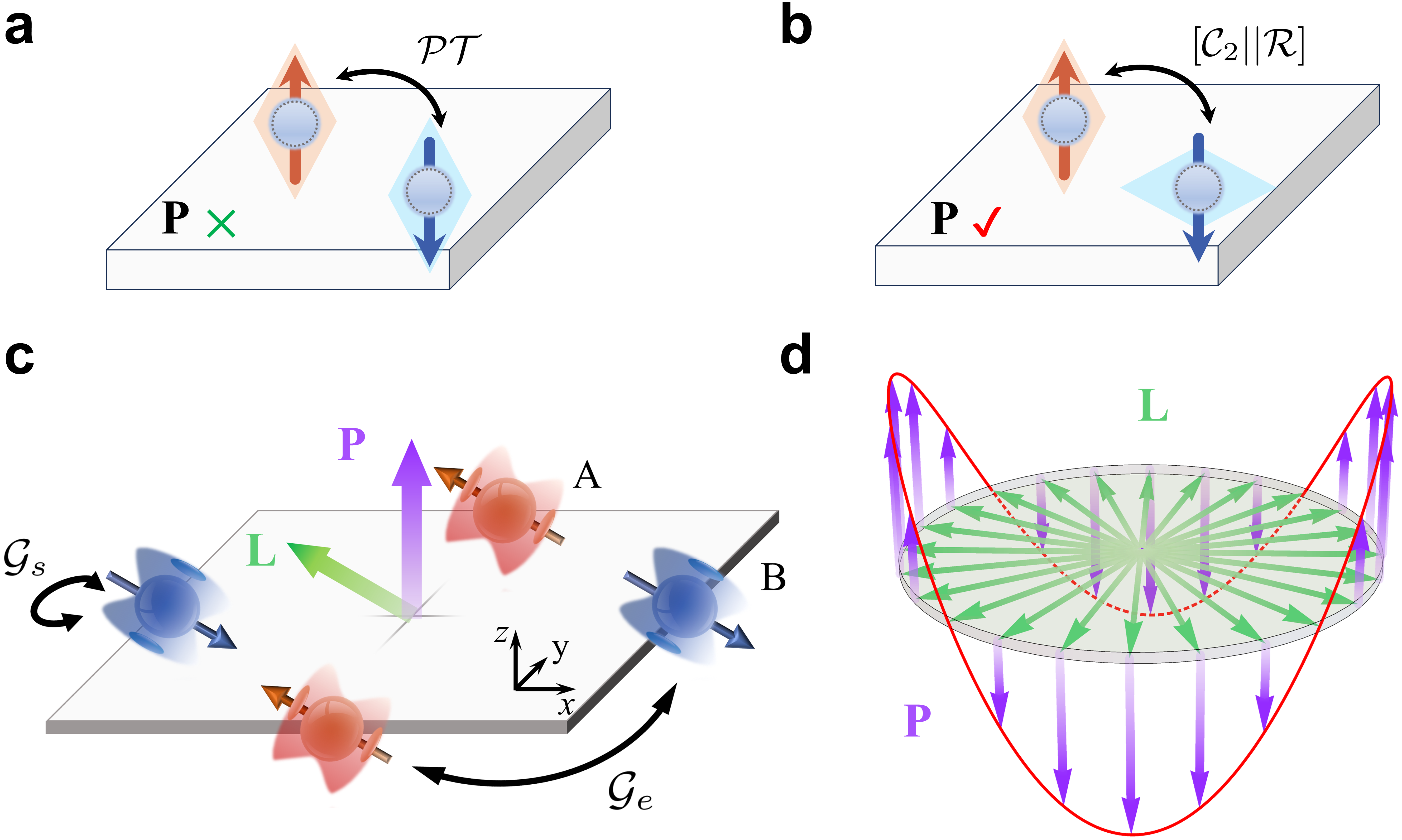}
		\par\end{centering}
	\caption{\textbf{Schematic of altermagnetism-induced multiferroics.} Illustration of electric polarization in conventional $\mathcal{PT}$-symmetric antiferromagnets (\textbf{a}) and altermagnets (\textbf{b}).  In \textbf{a}, the two opposite magnetic sublattices are related via $\mathcal{PT}$ symmetry, leading to compensated electric dipole moments and vanishing spontaneous electric polarization \textbf{P}. Whereas in \textbf{b}, the two sublattices are not connected by inversion, and thus finite \textbf{P} is generally permissible. \textbf{c}, Illustration of the locking behavior between \textbf{P} and N\'eel order \textbf{L} in altermagnets, which results from constraints imposed by $\mathcal{G}_s$ and $\mathcal{G}_e$ symmetries connecting the same and different sublattices, respectively. \textbf{d}, Typical locking behavior between out-of-plane \textbf{P} and in-plane \textbf{L} orientation for category 5 altermagnetic multiferroics in our classifcation.}
	\label{fig1}
\end{figure}

Multiferroicity, characterized by the coexistence of ferroelectricity and magnetism in a single-phase material, provides a unique perspective to investigate ME coupling mechanisms in magnetic systems. Multiferroic materials are broadly classified into type-I and type-II, with the latter demonstrating significantly stronger intrinsic ME coupling due to their magnetic-order-driven ferroelectricity. However, most experimentally reported type-II multiferroic materials have noncollinear magnetic order, whose complex response to external fields may significantly hinder the manipulation of ME effects~\cite{dong2019magnetoelectricity, li2023realistic, liu2024spin, stavric2025giant}. Therefore, achieving type-II multiferroicity in materials with collinear magnetic order is highly desirable. Interestingly, it has been proposed that ferroelectric polarization can emerge via metal-ligand hybridization in collinear spin systems, where crystal field effects break inversion symmetry at magnetic ion sites~\cite{murakawa2010ferroelectricity, zhang2022coexistence, zhou2024double}. Given that altermagnets inherently combine collinear magnetic order with broken inversion symmetry, a pivotal question emerges: can their unconventional spin structures induce electric polarization, thus establishing a  paradigm for type-II multiferroicity?

In this work, we employ symmetry analysis and metal-ligand theoretical modeling to definitively resolve the aforementioned issue by explicitly demonstrating the emergence of electric polarization induced by altermagnetic spin structures. Crucially, we establish a direct link between the electric polarization vector and the altermagnetic N\'{e}el order parameter, which unambiguously realizes type-II multiferroicity. Based on locking behaviors between electric polarization and N\'eel order orientation, we further classify two-dimensional (2D) altermagnetic type-II multiferroics governed by layer group (LG) symmetries into eight distinct categories. Then, we demonstrate the monolayer MgFe$_2$N$_2$ as a prototypical altermagnetic multiferroic material through first-principles calculations and robustly validate our theoretical framework. We also propose an experimental detection scheme via magneto-optical Faraday effect to identify the N\'eel order orientation and accompanying electric polarization in altermagnetic multiferroics. By combining strong ME couplings of type-II multiferroics and unique advantages of altermagnets, our work provides a versatile platform for designing multifunctional devices.\\

\noindent\textbf{Results}\\
\noindent\textbf{Altermagnetic type-II multiferroics.}
The electric dipole moment is odd under spatial inversion operation $\mathcal{P}$ but even under time-reversal operation $\mathcal{T}$. Consequently, in conventional $\mathcal{PT}$-symmetric antiferromagnets, where inversion symmetry connects the two antiparallel magnetic sublattices, the electric dipole moments from the two sublattices exactly cancel out, resulting in zero net polarization $\mathbf{P}$, as illustrated in Fig.~\ref{fig1}a. However, the symmetry constraint in altermagnets is fundamentally different. As antiparallel magnetic sublattices in altermagnets are 
never connected by inversion, their dipole moments no longer cancel out, permitting the emergence of a finite polarization, as shown in Fig.~\ref{fig1}b.

Next, we start from microscopic electric dipole moments to derive the expression of electric polarization induced by altermagnetic N\'{e}el order.  We consider an altermagnet with two antiparallel magnetic sublattices labeled as A and B, respectively [see Fig.~\ref{fig1}c], each hosting one magnetic ion per sublattice with spin $\mathbf{S}_{A(B)}=\pm \mathbf{S}$.  The N\'{e}el vector is defined as $\mathbf{L} = \frac{1}{2}(\mathbf{S}_{A} - \mathbf{S}_{B})= \mathbf{S}$. We further assume A and B sublattices occupy the same Wyckoff position of multiplicity two with the site symmetry group denoted as $\mathcal{G}_s$. In addition, for altermagnets, there always exists at least one real-space group symmetry operation connecting different sublattices, such as the fourfold rotation in Fig.~\ref{fig1}c, and we collect all such group operations into a set denoted as $\mathcal{G}_e$. According to the mechanism of spin-induced electronic polarization (e.g., metal-ligand hybridization mechanism)~\cite{murakawa2010ferroelectricity, jia2006bond, zhang2022coexistence, zhou2024double, matsumoto2017symmetry}, in the presence of spin-orbit coupling (SOC), the local electric dipole moment from each magnetic sublattice can be represented as a linear combination of products of spins: $p_{M}^{\alpha}=[K_M]_{\beta\gamma}^\alpha S_{M}^{\beta} S_{M}^{\gamma}$, where $p_{M}^{\alpha}$ and $S_{M}^{\alpha}$ ($\alpha = x, y, z,\ M=A,B$) are the $\alpha$-component  of the electric dipole moment $\mathbf{p}_M$  and spin $\mathbf{S}_M$ in Cartesian coordinates, respectively, and the coefficient $K_M$ is a real-symmetric third-rank polar tensor. Note that the existence of symmetry operations in $\mathcal{G}_s$($\mathcal{G}_e$) connecting the same (different) sublattices will impose constraints on the coefficient tensors [see Supplementary Note 1 for details], given by 
\begin{equation}
	[R_{s}]_{\alpha\beta} [K_{M}]^{\beta}_{\gamma\delta} = [R_{s}]^{T}_{\gamma\mu} [K_{M}]^{\alpha}_{\mu\nu} [R_{s}]_{\nu\delta},
\end{equation}
and
\begin{equation}
	[R_{e}]_{\alpha\beta} \left[K_{\overline{M}}\right]^{\beta}_{\gamma\delta} = [R_{e}]^T_{\gamma\mu} \left[K_{M}\right]^{\alpha}_{\mu\nu} [R_{e}]_{\nu\delta},
\end{equation}
where $R_s$ and $R_{e}$ are the point group parts of the symmetry operations in $\mathcal{G}_s$ and $\mathcal{G}_e$, respectively, and $\overline{M}$ denotes different sublattices. These constraints will significantly reduce independent elements of the coefficient tensors $K_A$ and $K_B$. In other words, the detailed form of the coefficient tensor is fully determined by the space group of the crystal and Wyckoff positions of the magnetic sublattices. In addition, the strength of SOC influences the magnitude of $K_M$, which typically increases as SOC strengthens (see Supplementary Note 4 for details).

The total electric dipole moment per unit cell $\mathbf{p}_\text{tot}$ can then be obtained by summing local electric dipole moments from the two sublattices, which can be expressed in the component form as
\begin{equation}
	\label{eq2}
	\textrm{p}_\text{tot}^\alpha=[p_A]^\alpha+[p_B]^\alpha=\left[K_A+K_B\right]_{\beta\gamma}^\alpha L^\beta L^\gamma,
\end{equation}
where $L^\beta$ is the $\beta$-component of N\'{e}el vector $\mathbf{L}$. For an altermagnet containing $N$ unit cells, the macroscopic electric polarization is given by $\mathbf{P}=N\mathbf{p}_\text{tot}$, demonstrating a direct correlation between polarization and N\'{e}el vector. For conventional $\mathcal{PT}$-symmetric antiferromagnets, according to Eq.~(2), the inversion operation $\mathcal{P}=-\delta_{\alpha\beta}\ (\alpha,\beta=x,y,z)$ enforces $K_{A}=-K_{B}$, consequently yielding $\mathbf{P}=0$. In contrast, altermagnets intrinsically violate this symmetry condition, typically exhibiting $K_{A}\neq-K_{B}$ due to the lack of inversion operation. Notably, for altermagnets whose nonmagnetic state is characterized by a nonpolar point group with vanishing polarization, the N\'{e}el order can change it to a magnetic state featuring a polar magnetic point group and nonzero polarization. In this case, the electric polarization is directly induced by the magnetic order, thereby realizing type-II multiferroics in altermagnets.

Based on the above discussion, altermagnetic type-II multiferroics are enabled by the symmetry conditions and magnetic structure inherent to altermagnetism, while SOC establishes the microscopic mechanism for the magnetic-order-induced electric polarization. Although SOC is typically not large for altermagnets, it usually plays a significant role in various phenomena of altermagnets, such as the anomalous Hall effect. Therefore, type-II multiferroics provide another prominent example of how the interplay between SOC and altermagnetism gives rise to rich physical effects. Moreover, owing to the underlying mechanism of type-II multiferroics, substantial ME couplings can be achieved without requiring a large SOC strength. It is also worth mentioning that, while some traditional collinear compensated magnets, specifically the $T+\tau$ ($\tau$ is fractional translation) antiferromagnets with broken $PT$ symmetry can in principle host type-II multiferroics, altermagnets offer unique and significant advantages (see Supplementary Note 5 for detailed comparison). Altermagnets not only possess significantly larger spin splittings, but also exhibit far richer ME coupling behaviors,  which could offer better tunability for applications of type-II multiferroics.\\

\begin{table}[!htbp]
\centering
\renewcommand{\arraystretch}{1.15}
\caption{\textbf{Layer-group classification of 2D altermagnetic type-II multiferroics.} The layer groups satisfying altermagnetic type-II multiferroics are classified into eight categories (Cat.), distinguished by distinct locking behaviors between the polarization $\mathbf{P}$ and the N\'{e}el vector ($\mathbf{L}$) orientation, with $\theta$ ($\phi$) denoting the polar (azimuthal) angle of $\mathbf{L}$ in spherical coordinates. The Wyckoff position and corresponding site symmetry (Site symm.) of the magnetic sublattice are also presented. Here, the parameters $A=\alpha \sin(2\theta), B=\beta \sin^2\theta, C=\gamma \sin(2\theta), D=\delta \sin^2\theta$, where $\alpha, \beta, \gamma, \delta$ are material-dependent constants.}
\label{classification}

\resizebox{\columnwidth}{!}{%
\begin{tabular}{c c c c c}
\specialrule{0.5pt}{0pt}{0pt}
\specialrule{0.5pt}{-2pt}{3pt}
Cat. & LG & Wyckoff & Site symm. & Polarization\Tstrut\Bstrut \\
\midrule

\catcelltwo{1} & 19 & 2l,2k,2j,2i & ..2 &
\polbox{A\sin\phi\\ C\cos\phi\\ B\sin(2\phi)}
\Tstrut\Bstrut \\
& 20,21 & 2b,2a & ..2 &
\polbox{A\sin\phi\\ C\cos\phi\\ B\sin(2\phi)}
\Tstrut\Bstrut \\
\midrule

2 & 50 & 2e,2d,2c & 2.. &
\polbox{A\cos\phi + C\sin\phi\\ C\cos\phi - A\sin\phi\\ B\cos(2\phi)+D\sin(2\phi)}
\Tstrut\Bstrut \\
\midrule

\catcellthree{3} & 53 & 2e,2d(2c) & 4..(222.) &
\polbox{A\sin\phi\\ -A\cos\phi\\ 0}
\Tstrut\Bstrut \\
& 54 & 2b(2a) & 4..(2.22) &
\polbox{A\sin\phi\\ -A\cos\phi\\ 0}
\Tstrut\Bstrut \\
& 76 & 2c & 6.. &
\polbox{A\sin\phi\\ -A\cos\phi\\ 0}
\Tstrut\Bstrut \\
\midrule

\catcelltwo{4} & 57 & 2e,2d(2c) & 2.mm(222.) &
\polbox{A\sin\phi\\ A\cos\phi\\ B\sin(2\phi)}
\Tstrut\Bstrut \\
& 58 & 2b(2a) & 2.mm(-4..) &
\polbox{A\sin\phi\\ A\cos\phi\\ B\sin(2\phi)}
\Tstrut\Bstrut \\
\midrule

\catcelltwo{5} & 59 & 2e,2d,2c & 2mm. &
\polbox{A\cos\phi\\ -A\sin\phi\\ B\cos(2\phi)}
\Tstrut\Bstrut \\
& 60 & 2b(2a) & 2.22(-4..) &
\polbox{A\cos\phi\\ -A\sin\phi\\ B\cos(2\phi)}
\Tstrut\Bstrut \\
\midrule

6 & 67 & 2f,2e,2d & 3.. &
\polbox{A\sin\phi + B\sin(2\phi)\\ -A\cos\phi + B\cos(2\phi)\\ 0}
\Tstrut\Bstrut \\
\midrule

7 & 68 & 2c,2b & 3.. &
\polbox{A\sin\phi + B\cos(2\phi)\\ -A\cos\phi - B\sin(2\phi)\\ 0}
\Tstrut\Bstrut \\
\midrule

8 & 79 & 2b & -6.. &
\polbox{B\cos(2\phi)\\ -B\sin(2\phi)\\ 0}
\Tstrut\Bstrut \\
\specialrule{0.5pt}{2pt}{0pt}
\specialrule{0.5pt}{-2pt}{0pt}
\end{tabular}%
}
\end{table}

\noindent\textbf{Classification of 2D altermagnetic type-II multiferroics.}~Atomically thin 2D multiferroic materials exhibit unique structural and functional merits, such as enhanced ME coupling, ultralow energy dissipation, and multifunctional integration capabilities, positioning them as ideal candidates for next-generation nanoelectronics and flexible device applications~\cite{wu2024coexistence, matsukura2015control, jiang2018controlling, song2022evidence}. Focusing on type-II multiferroics in 2D altermagnetic systems whose crystal symmetry is described by layer groups (LGs), we propose a four-step classification scheme as follows. First, we comprehensively analyze all 80 LGs to identify those containing multiplicity-two Wyckoff positions that satisfy the symmetry requirements for altermagnetic systems with two magnetic sublattices. Second, we filter out the polar LGs belonging to type-I multiferroics (see Supplementary Table 1 for more details). Third, for the remaining nonpolar LGs, we exclude those where the site symmetry group of multiplicity-two Wyckoff positions includes inversion symmetry. Fourth, we derive explicit expressions of electric polarization $\mathbf{P}$ through Eqs.~(1-3) as a function of N\'{e}el vector $\mathbf{L}$ for each LG.

Based on distinct locking configurations between polarization and the N\'{e}el vector orientation, we classify type-II multiferroics in 2D altermagnets into eight categories in terms of LGs, as presented in Table I, where $\theta$ and $\phi$ are the polar and azimuthal angles of $\mathbf{L}$ in spherical coordinates (see Supplementary Note 2 for details). Notably, each component of $\mathbf{P}$ exhibits either $\pi$- or $2\pi$-periodic behaviors with respect to azimuthal angle $\phi$ of $\mathbf{L}$ for all categories. To exemplify this classification, we select category 5 as a representative case where $\mathbf{L}$ lies in the $x$-$y$ plane. In this case, the polarization is always out-of-plane with $\mathrm{P}\propto\cos (2\phi)$, showing a $\pi$-periodic dependence on the azimuthal angle of $\mathbf{L}$ as illustrated in Fig.~\ref{fig1}d. For completeness, we also performed a similar space-group classification for three-dimensional altermagnetic type-II multiferroics (see Supplementary Table 2 for details). \\

\begin{figure}[tbp]
	\begin{centering}
		\includegraphics[width=0.45\textwidth]{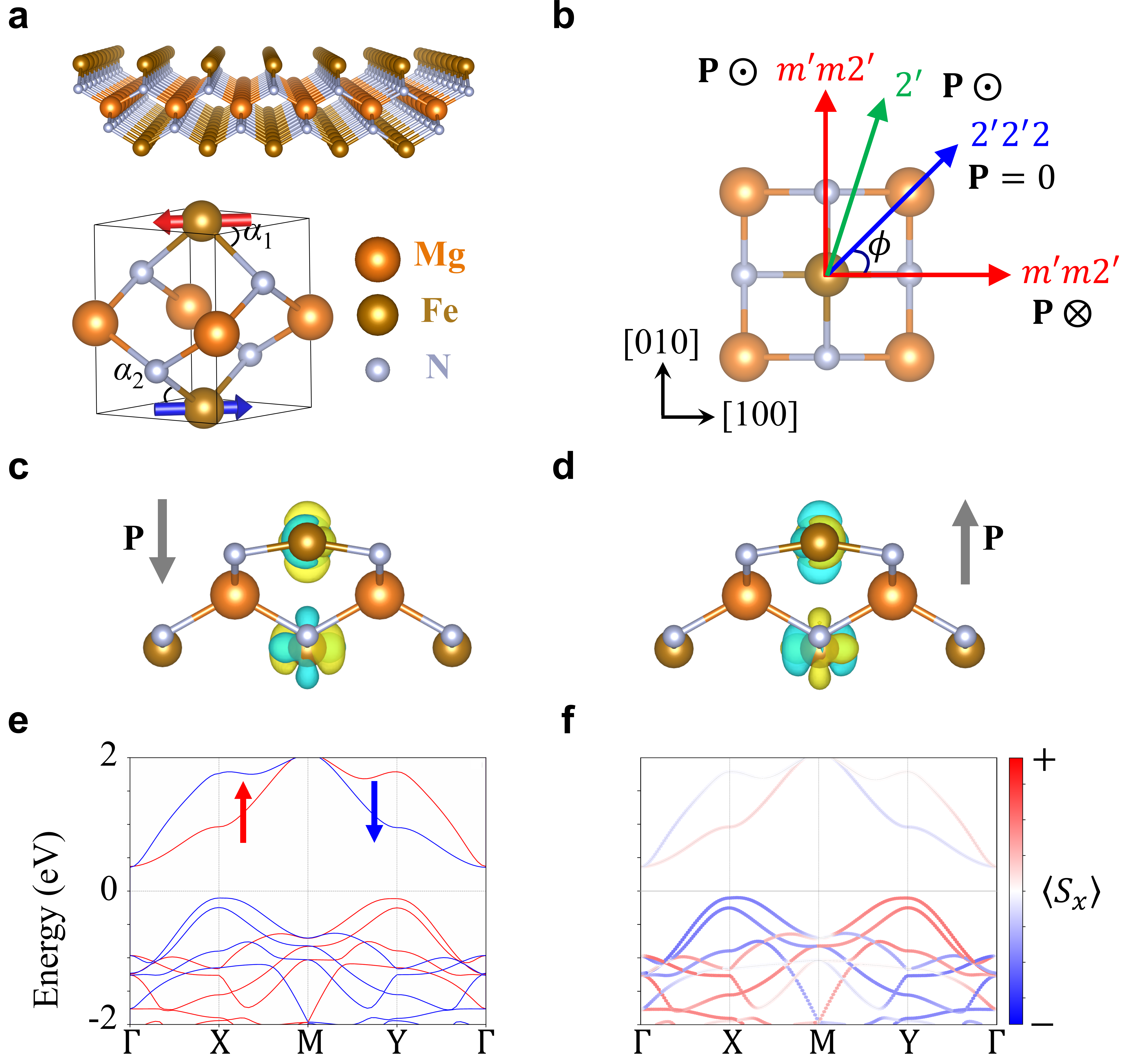}
		\par\end{centering}
	\caption{\textbf{Type-II multiferroicity in altermagnetic monolayer MgFe$_2$N$_2$.} Crystal structure of monolayer MgFe$_2$N$_2$ shown in front and side views (\textbf{a}), and top view (\textbf{b}). The in-plane N\'eel order originates from Fe atoms. Magnetic point group symmetries and associated electric polarization for distinct N\'eel order orientations are summarized in \textbf{b}. The magnetic-order-induced charge density difference (compared to the paraelectric phase) around Fe atoms for the inverse ferroelectric ($P_z<0$)  (\textbf{c}) and ferroelectric phases ($P_z>0$) (\textbf{d}) at $\phi=0$ and $\phi=\pi/2$, respectively, where distinct charge transfer trends can be seen. Spin-resolved band structures of monolayer MgFe$_2$N$_2$ without (\textbf{e}) and with (\textbf{f}) spin-orbit coupling, respectively, where the N\'eel vector is aligned along the crystallographic [100]-direction.}
	\label{fig2}
\end{figure}

\noindent\textbf{Altermagnetic type-II multiferroicity in monolayer MgFe$_2$N$_2$.}~To validate the preceding theoretical framework, we propose monolayer MgFe$_2$N$_2$ as a representative material exhibiting altermagnetic type-II multiferroicity. The nonmagnetic state of monolayer MgFe$_2$N$_2$ [see Figs.~\ref{fig2}(a,b)] crystallizes in  LG No.~59, featuring nonpolar $\bar{4}$2$m$ point group symmetry. This indicates the absence of ferroelectricity in nonmagnetic monolayer MgFe$_2$N$_2$. Our first-principles calculations and symmetry analysis identify altermagnetic ordering characterized by in-plane antiparallel alignment of magnetic moments from Fe atoms [see Fig.~\ref{fig2}a] as the magnetic ground state of monolayer MgFe$_2$N$_2$, which is energetically favorable over the in-plane parallel alignment by at least approximately 70\textrm{~meV} f.u.$^{-1}$~(see Supplementary  3 for the total energy calculation). The band structures without [Fig.~\ref{fig2}e] and with SOC [Fig.~\ref{fig2}f] demonstrate nearly identical energy spectra accompanied by pronounced momentum-dependent spin splittings (reaching 1 eV at some $k$ points). This substantiates the altermagnetic state of monolayer MgFe$_2$N$_2$, where SOC exerts negligible effects in energy spectra and magnetic order.

\begin{figure}[tbp]
	\begin{centering}
		\includegraphics[width=0.45\textwidth]{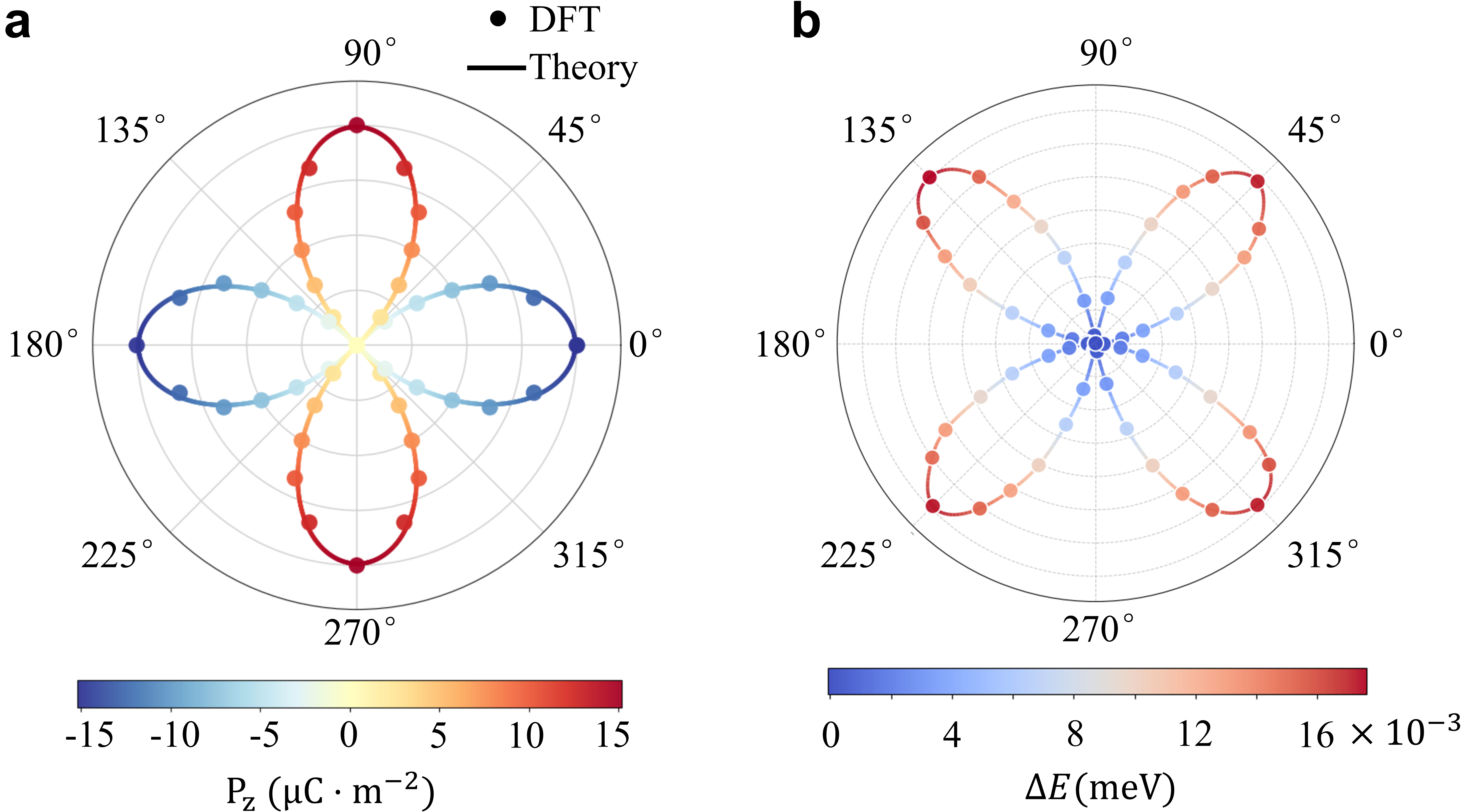}
		\par\end{centering}
	\caption{\textbf{N\'eel-order-locked electric polarization of altermagnetic monolayer MgFe$_2$N$_2$}. First-principles calculations of out-of-plane electric polarization $P_z$ (\textbf{a}) and total energy difference compared to the minimum energy (\textbf{b}) as a function of in-plane N\'eel order orientation. Here, $\phi=0^{\circ}$ corresponds to the crystallographic [100] direction. }
	\label{fig3}
\end{figure}

Monolayer MgFe$_2$N$_2$ belongs to category 5 in our classification of 2D altermagnetic type-II multiferroics. Since its ground state N\'eel order lies in the $x$-$y$ plane, the electric polarization is expected to have only out-of-plane component $\mathrm{P}_z$ which is locked with the polar angle $\phi$ of the in-plane N\'eel order as $\mathrm{P}_z\propto\cos (2\phi)$. This is confirmed by our first-principles calculations shown in Fig.~\ref{fig3}a, where $\mathrm{P}_z$ exhibits a $\pi$-periodic behavior of $\phi$ (the crystallographic [100]-direction is set as $\phi=0$). Specifically, $\mathrm{P}_z$ reaches maximal values of $\pm15.2~\mu\text{C}\cdot\text{m}^{-2}$ at $\phi=n\pi/2$ $(n=0,1,2,3)$ and vanishes at $\phi=(2n+1)\pi/4$. Note that with varying $\phi$ from 0 to $\pi/2$ (or $\pi$ to $3\pi/2$), there exists a sign reversal of the maximal $P_z$ value from the inverse ferroelectric phase ($P_z<0$) at $\phi=0$ to the ferroelectric phase  ($P_z>0$)  at $\phi=\pi/2$. This polarization switching process only requires overcoming a negligibly small energy barrier below 0.02 meV [See Supplementary Note 3 for the results of Fe$_2$WTe$_4$], as shown in Fig.~\ref{fig3}b, thus demonstrating the potential for efficient ME control of polarization through N\'eel order manipulation. In fact, the polarization switching process can also be understood by magnetic point group analysis. As illustrated in Fig.~\ref{fig2}b, during the in-plane rotation of $\mathbf{L}$, the system transitions sequentially from the polar magnetic point group $m'm$2$'$ at $\phi=0$ to the nonpolar phase 2$'$2$'$2 at $\phi=\pi/4$ ($\mathrm{P}_z=0$) via an intermediate polar phase 2$'$ at other angles, eventually recovering $m'm$2$'$ symmetry at $\phi=\pi/2$. Note that the reversal of $P_z$ upon a $\pi/2$ change of $\phi$ can be understood from the four-fold rotoinversion symmetry $S_{4z}$ of the MgFe$_2$N$_2$ lattice, which ensures that a $C_{4z}$ rotation of the lattice maps it onto its $M_z$-linked counterpart, with a flip of $P_z$ (see Supplementary Note 3).

Based on the metal-ligand hybridization mechanism~\cite{murakawa2010ferroelectricity, murakawa2012comprehensive, zhou2024double, matsumoto2017symmetry, jia2006bond, jia2007Microscopic} for type-II multiferroics, we give a microscopic explanation for altermagnetic multiferroicity in monolayer MgFe$_2$N$_2$ as follows. As the local crystal field of MgFe$_2$N$_2$ breaks inversion symmetry [see Fig.~\ref{fig2}a], the even-parity $d$ orbitals of Fe ions mix up with the odd-parity $p$ orbitals of the ligand N ions, inducing the non-zero electric dipole $\mathbf{p}\propto \sum_{i}(\mathbf{S} \cdot \hat{\mathbf{e}}_i)^2\hat{\mathbf{e}}_i$, where $\mathbf{S}$ represents the spin of Fe ion, while $\hat{\mathbf{e}}_i$ denotes the unit vector along the $i$-th Fe-N bond. As a result, the polarization in MgFe$_2$N$_2$ directly originates from its magnetic structures, demonstrating robust intrinsic ME coupling. For in-plane spins $\mathbf{S} = (S_x, S_y, 0) \propto (\cos\phi, \sin\phi, 0)$, the resulting net electric dipole can be obtained as $\mathrm{p}_z \propto \cos(2\phi)$, which is consistent with the result from LG-based classification. As a further evidence, we have plotted the magnetic-order-induced charge density difference around Fe ions for $\phi=0$ and $\phi=\pi/2$ in Figs.~\ref{fig2}c and~\ref{fig2}d, respectively. Distinct charge transfer trends can be seen between the two (inverse)ferroelectric phases, serving as a direct signature of the polarization reversal. \\

\begin{figure}[tbp]
	\begin{centering}
		\includegraphics[width=0.45\textwidth]{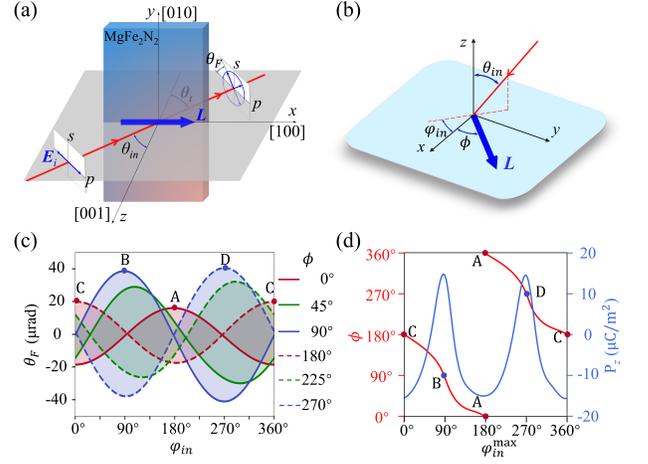}
		\par\end{centering}
	\caption{\textbf{Detection of N\'eel vector orientation via magneto-optical microscopy for altermagnetic monolayer MgFe$_2$N$_2$.} \textbf{a}, Schematic of the magneto-optical Faraday measurement, where a $p$-polarized light beam is incident on monolayer MgFe$_2$N$_2$ along the direction characterized by the polar ($\theta_{in}$) and azimuthal ($\varphi_{in}$) angles in spherical coordinates. $\theta_t$ and $\theta_{F}$ denote the refraction and the Faraday angles, respectively.  \textbf{b}, Illustration of $\varphi_{in}$, $\theta_{in}$ and $\phi$. \textbf{c}, $\theta_{F}$ versus $\varphi_{in}$ at fixed photon energy $\hbar\omega=0.1$ eV and $\theta_{in} = 60^{\circ}$ for representative \textbf{L} orientations, all of which exhibit $2\pi$-periodic trigonometric dependencies. \textbf{d}, The correspondence between $\varphi_{in}^{\textrm{max}}$ [where maximum $\theta_{F}$ is achieved, as marked in \textbf{c})] and in-plane N\'eel order orientation $\phi$ (red lines) and out-of-plane polarization $P_z$ (blue lines) with $\hbar\omega=0.1$ eV and $\theta_{in} = 60^{\circ}$.}
	\label{fig4}
\end{figure}

\noindent\textbf{Detection and manipulation of magnetoelectric couplings.}~Given the intrinsic coupling between polarization and N\'eel order in altermagnetic multiferroics, identifying and controlling the N\'eel vector orientation is therefore essential for probing and manipulating magnetoelectric effects. Here, we still take MgFe$_2$N$_2$ monolayer as a representative material to show that in-plane N\'eel vector orientation could be unambiguously determined via measurements of Faraday angles. As illustrated in Figs.~\ref{fig4}a and \ref{fig4}b, we consider a $p$-polarized light beam incident on monolayer MgFe$_2$N$_2$ at an angle $\theta_{in}$ with respect to the $z$-axis and an azimuthal angle $\varphi_{in}$ relative to the $x$-axis. Note that $\varphi_{in}$ can be easily tuned between $0$ and $2\pi$ by rotating the light beam around the $z$-direction. The calculated Faraday angle $\theta_F$ as a function of $\varphi_{in}$ at fixed $\theta_{in}=60^\circ$ and photon energy of $0.1$ eV for representative N\'eel vector orientations are presented in Fig.~\ref{fig4}c, where $\theta_F$ exhibits a $2\pi$-periodic trigonometric dependence on $\varphi_{in}$. We find a specific one-to-one relationship between the N\'eel vector orientation ($\phi$) and the critical azimuthal angle ($\varphi_{in}^{\text{max}}$) displaying the maximum $\theta_F$ value, which provides a feasible method to determine the N\'eel vector orientation [see  Supplementary Note 6 for details], as demonstrated in Fig.~\ref{fig4}d for $\theta_{in}=60^\circ$. Furthermore, we also propose an experimental scheme by leveraging the ultrafast N\'eel spin--orbit torque technique~\cite{behovits2023Terahertz} to control the N\'eel vector and thereby manipulate the polarization and ME couplings (see Supplementary Note 7 for details). \\\\

\noindent\textbf{Discussion}\\
We propose a paradigm for realizing type-II multiferroicity in altermagnets whose ferroelectric polarization is generated by unconventional spin structures. By establishing the correlation between the N\'eel vector and emergent polarization via spin-dependent dipole-moment analysis, we demonstrate symmetry-dependent locking behaviors, which are systematically classified into eight distinct categories based on LG symmetries in 2D altermagnets. First-principles calculations on monolayer MgFe$_2$N$_2$ validate this mechanism, revealing spin-orientation-dependent polarization switching. We also propose a detection scheme based on magneto-optical Faraday effect to identify the orientation of both the N\'eel vector and its coupled electric polarization. This work provides a foundation for exploring magnetoelectric couplings in altermagnets and enables advancements in low-energy-consumption spintronic devices via symmetry-guided material design.

\noindent\textbf{Note added.}~We were aware that recent studies have also focused on the ferroelectric properties of altermagnets~\cite{duan2025antiferroelectric, gu2025ferroelectric, sun2025proposing, vsmejkal2024altermagnetic, cao2024designing, zhu2025two, zhu2025sliding, zhu2025emergent,guo2025Mechanically,Urru2025gtype}.\\

\noindent\textbf{Acknowledgements}\\
This work is supported by the National Key Projects for Research and Development of China (Grant Nos. 2024YFA1409100 to H.Z., 2021YFA1400400 to H.Z.), the Natural Science Foundation of China (Grant Nos. 12534007 to H.Z. and H.W., 92365203 to H.Z.), the Natural Science Foundation of Jiangsu Province (Grant Nos. BK20253012 to H.Z., BK20252117 to H.W., BK20243011 to H.Z., BK20233001 to H.Z.), the Fundamental Research Funds for the Central Universities (Grant No. KG202501 to H.Z.), the Fundamental and Interdisciplinary Disciplines Breakthrough Plan of the Ministry of Education of China (Grant No. JYB2025XDXM409 to H.Z.), the China Postdoctoral Science Foundation (Grant No. 2024M761372 to W.G.), the Postdoctoral Fellowship Program Grade C of China Postdoctoral Science Foundation (Grant No. GZC20252201 to W.G.), and the e-Science Center of Collaborative Innovation Center of Advanced Microstructures.\\


\bibliography{ref} 

\begin{thebibliography}{76}%
\makeatletter
\providecommand \@ifxundefined [1]{%
 \@ifx{#1\undefined}
}%
\providecommand \@ifnum [1]{%
 \ifnum #1\expandafter \@firstoftwo
 \else \expandafter \@secondoftwo
 \fi
}%
\providecommand \@ifx [1]{%
 \ifx #1\expandafter \@firstoftwo
 \else \expandafter \@secondoftwo
 \fi
}%
\providecommand \natexlab [1]{#1}%
\providecommand \enquote  [1]{``#1''}%
\providecommand \bibnamefont  [1]{#1}%
\providecommand \bibfnamefont [1]{#1}%
\providecommand \citenamefont [1]{#1}%
\providecommand \href@noop [0]{\@secondoftwo}%
\providecommand \href [0]{\begingroup \@sanitize@url \@href}%
\providecommand \@href[1]{\@@startlink{#1}\@@href}%
\providecommand \@@href[1]{\endgroup#1\@@endlink}%
\providecommand \@sanitize@url [0]{\catcode `\\12\catcode `\$12\catcode
  `\&12\catcode `\#12\catcode `\^12\catcode `\_12\catcode `\%12\relax}%
\providecommand \@@startlink[1]{}%
\providecommand \@@endlink[0]{}%
\providecommand \url  [0]{\begingroup\@sanitize@url \@url }%
\providecommand \@url [1]{\endgroup\@href {#1}{\urlprefix }}%
\providecommand \urlprefix  [0]{URL }%
\providecommand \Eprint [0]{\href }%
\providecommand \doibase [0]{https://doi.org/}%
\providecommand \selectlanguage [0]{\@gobble}%
\providecommand \bibinfo  [0]{\@secondoftwo}%
\providecommand \bibfield  [0]{\@secondoftwo}%
\providecommand \translation [1]{[#1]}%
\providecommand \BibitemOpen [0]{}%
\providecommand \bibitemStop [0]{}%
\providecommand \bibitemNoStop [0]{.\EOS\space}%
\providecommand \EOS [0]{\spacefactor3000\relax}%
\providecommand \BibitemShut  [1]{\csname bibitem#1\endcsname}%
\let\auto@bib@innerbib\@empty
\bibitem [{\citenamefont {Ma}\ \emph {et~al.}(2021)\citenamefont {Ma},
  \citenamefont {Hu}, \citenamefont {Li}, \citenamefont {Liu}, \citenamefont
  {Yao}, \citenamefont {Jia},\ and\ \citenamefont
  {Liu}}]{ma2021aMultifunctiona}%
  \BibitemOpen
  \bibfield  {author} {\bibinfo {author} {\bibfnamefont {H.-Y.}\ \bibnamefont
  {Ma}}, \bibinfo {author} {\bibfnamefont {M.}~\bibnamefont {Hu}}, \bibinfo
  {author} {\bibfnamefont {N.}~\bibnamefont {Li}}, \bibinfo {author}
  {\bibfnamefont {J.}~\bibnamefont {Liu}}, \bibinfo {author} {\bibfnamefont
  {W.}~\bibnamefont {Yao}}, \bibinfo {author} {\bibfnamefont {J.-F.}\
  \bibnamefont {Jia}},\ and\ \bibinfo {author} {\bibfnamefont {J.}~\bibnamefont
  {Liu}},\ }\bibfield  {title} {\bibinfo {title} {Multifunctional
  antiferromagnetic materials with giant piezomagnetism and noncollinear spin
  current},\ }\href {https://doi.org/10.1038/s41467-021-23127-7} {\bibfield
  {journal} {\bibinfo  {journal} {Nat. Commun.}\ }\textbf {\bibinfo {volume}
  {12}},\ \bibinfo {pages} {2846} (\bibinfo {year} {2021})}\BibitemShut
  {NoStop}%
\bibitem [{\citenamefont {Mazin}\ \emph {et~al.}(2021)\citenamefont {Mazin},
  \citenamefont {Koepernik}, \citenamefont {Johannes}, \citenamefont
  {Gonz{\'a}lez-Hern{\'a}ndez},\ and\ \citenamefont
  {{\v{S}}mejkal}}]{mazin2021prediction}%
  \BibitemOpen
  \bibfield  {author} {\bibinfo {author} {\bibfnamefont {I.~I.}\ \bibnamefont
  {Mazin}}, \bibinfo {author} {\bibfnamefont {K.}~\bibnamefont {Koepernik}},
  \bibinfo {author} {\bibfnamefont {M.~D.}\ \bibnamefont {Johannes}}, \bibinfo
  {author} {\bibfnamefont {R.}~\bibnamefont {Gonz{\'a}lez-Hern{\'a}ndez}},\
  and\ \bibinfo {author} {\bibfnamefont {L.}~\bibnamefont {{\v{S}}mejkal}},\
  }\bibfield  {title} {\bibinfo {title} {{Prediction of unconventional
  magnetism in doped FeSb$_2$}},\ }\href
  {https://doi.org/10.1073/pnas.2108924118} {\bibfield  {journal} {\bibinfo
  {journal} {Proc. Natl. Acad. Sci.}\ }\textbf {\bibinfo {volume} {118}},\
  \bibinfo {pages} {e2108924118} (\bibinfo {year} {2021})}\BibitemShut
  {NoStop}%
\bibitem [{\citenamefont {Gonz{\'a}lez-Hern{\'a}ndez}\ \emph
  {et~al.}(2021)\citenamefont {Gonz{\'a}lez-Hern{\'a}ndez}, \citenamefont
  {{\v{S}}mejkal}, \citenamefont {V{\`y}born{\`y}}, \citenamefont {Yahagi},
  \citenamefont {Sinova}, \citenamefont {Jungwirth},\ and\ \citenamefont
  {{\v{Z}}elezn{\`y}}}]{gonzalez2021efficient}%
  \BibitemOpen
  \bibfield  {author} {\bibinfo {author} {\bibfnamefont {R.}~\bibnamefont
  {Gonz{\'a}lez-Hern{\'a}ndez}}, \bibinfo {author} {\bibfnamefont
  {L.}~\bibnamefont {{\v{S}}mejkal}}, \bibinfo {author} {\bibfnamefont
  {K.}~\bibnamefont {V{\`y}born{\`y}}}, \bibinfo {author} {\bibfnamefont
  {Y.}~\bibnamefont {Yahagi}}, \bibinfo {author} {\bibfnamefont
  {J.}~\bibnamefont {Sinova}}, \bibinfo {author} {\bibfnamefont
  {T.}~\bibnamefont {Jungwirth}},\ and\ \bibinfo {author} {\bibfnamefont
  {J.}~\bibnamefont {{\v{Z}}elezn{\`y}}},\ }\bibfield  {title} {\bibinfo
  {title} {Efficient electrical spin splitter based on nonrelativistic
  collinear antiferromagnetism},\ }\href
  {https://doi.org/10.1103/PhysRevLett.126.127701} {\bibfield  {journal}
  {\bibinfo  {journal} {Phys. Rev. Lett.}\ }\textbf {\bibinfo {volume} {126}},\
  \bibinfo {pages} {127701} (\bibinfo {year} {2021})}\BibitemShut {NoStop}%
\bibitem [{\citenamefont {{{\v S}mejkal, Libor and Sinova, Jairo and Jungwirth,
  Tomas}}(2022)}]{smejkal2022emerging}%
  \BibitemOpen
  \bibfield  {author} {\bibinfo {author} {\bibnamefont {{{\v S}mejkal, Libor
  and Sinova, Jairo and Jungwirth, Tomas}}},\ }\bibfield  {title} {\bibinfo
  {title} {Emerging research landscape of altermagnetism},\ }\href
  {https://doi.org/10.1103/PhysRevX.12.040501} {\bibfield  {journal} {\bibinfo
  {journal} {Phys. Rev. X}\ }\textbf {\bibinfo {volume} {12}},\ \bibinfo
  {pages} {040501} (\bibinfo {year} {2022})}\BibitemShut {NoStop}%
\bibitem [{\citenamefont {{\v S}mejkal}\ \emph
  {et~al.}(2022{\natexlab{a}})\citenamefont {{\v S}mejkal}, \citenamefont
  {Sinova},\ and\ \citenamefont {Jungwirth}}]{smejkal2022conventional}%
  \BibitemOpen
  \bibfield  {author} {\bibinfo {author} {\bibfnamefont {L.}~\bibnamefont {{\v
  S}mejkal}}, \bibinfo {author} {\bibfnamefont {J.}~\bibnamefont {Sinova}},\
  and\ \bibinfo {author} {\bibfnamefont {T.}~\bibnamefont {Jungwirth}},\
  }\bibfield  {title} {\bibinfo {title} {Beyond conventional ferromagnetism and
  antiferromagnetism: A phase with nonrelativistic spin and crystal rotation
  symmetry},\ }\href {https://doi.org/10.1103/PhysRevX.12.031042} {\bibfield
  {journal} {\bibinfo  {journal} {Phys. Rev. X}\ }\textbf {\bibinfo {volume}
  {12}},\ \bibinfo {pages} {031042} (\bibinfo {year}
  {2022}{\natexlab{a}})}\BibitemShut {NoStop}%
\bibitem [{\citenamefont {Turek}(2022)}]{turek2022altermagnetism}%
  \BibitemOpen
  \bibfield  {author} {\bibinfo {author} {\bibfnamefont {I.}~\bibnamefont
  {Turek}},\ }\bibfield  {title} {\bibinfo {title} {Altermagnetism and magnetic
  groups with pseudoscalar electron spin},\ }\href
  {https://doi.org/10.1103/PhysRevB.106.094432} {\bibfield  {journal} {\bibinfo
   {journal} {Phys. Rev. B}\ }\textbf {\bibinfo {volume} {106}},\ \bibinfo
  {pages} {094432} (\bibinfo {year} {2022})}\BibitemShut {NoStop}%
\bibitem [{\citenamefont {{\v S}mejkal}\ \emph
  {et~al.}(2022{\natexlab{b}})\citenamefont {{\v S}mejkal}, \citenamefont
  {MacDonald}, \citenamefont {Sinova}, \citenamefont {Nakatsuji},\ and\
  \citenamefont {Jungwirth}}]{smejkal2022anomalous}%
  \BibitemOpen
  \bibfield  {author} {\bibinfo {author} {\bibfnamefont {L.}~\bibnamefont {{\v
  S}mejkal}}, \bibinfo {author} {\bibfnamefont {A.~H.}\ \bibnamefont
  {MacDonald}}, \bibinfo {author} {\bibfnamefont {J.}~\bibnamefont {Sinova}},
  \bibinfo {author} {\bibfnamefont {S.}~\bibnamefont {Nakatsuji}},\ and\
  \bibinfo {author} {\bibfnamefont {T.}~\bibnamefont {Jungwirth}},\ }\bibfield
  {title} {\bibinfo {title} {Anomalous hall antiferromagnets},\ }\href
  {https://doi.org/10.1038/s41578-022-00430-3} {\bibfield  {journal} {\bibinfo
  {journal} {Nat. Rev. Mater.}\ }\textbf {\bibinfo {volume} {7}},\ \bibinfo
  {pages} {482} (\bibinfo {year} {2022}{\natexlab{b}})}\BibitemShut {NoStop}%
\bibitem [{\citenamefont {Liu}\ \emph {et~al.}(2022)\citenamefont {Liu},
  \citenamefont {Li}, \citenamefont {Han}, \citenamefont {Wan},\ and\
  \citenamefont {Liu}}]{liu2022spin-group}%
  \BibitemOpen
  \bibfield  {author} {\bibinfo {author} {\bibfnamefont {P.}~\bibnamefont
  {Liu}}, \bibinfo {author} {\bibfnamefont {J.}~\bibnamefont {Li}}, \bibinfo
  {author} {\bibfnamefont {J.}~\bibnamefont {Han}}, \bibinfo {author}
  {\bibfnamefont {X.}~\bibnamefont {Wan}},\ and\ \bibinfo {author}
  {\bibfnamefont {Q.}~\bibnamefont {Liu}},\ }\bibfield  {title} {\bibinfo
  {title} {Spin-group symmetry in magnetic materials with negligible spin-orbit
  coupling},\ }\href {https://doi.org/10.1103/PhysRevX.12.021016} {\bibfield
  {journal} {\bibinfo  {journal} {Phys. Rev. X}\ }\textbf {\bibinfo {volume}
  {12}},\ \bibinfo {pages} {021016} (\bibinfo {year} {2022})}\BibitemShut
  {NoStop}%
\bibitem [{\citenamefont {{Krempask{\'y}, J. and {\v S}mejkal, L. and D'Souza,
  S. W. and others}}(2024)}]{krempasky2024altermagnetic}%
  \BibitemOpen
  \bibfield  {author} {\bibinfo {author} {\bibnamefont {{Krempask{\'y}, J. and
  {\v S}mejkal, L. and D'Souza, S. W. and others}}},\ }\bibfield  {title}
  {\bibinfo {title} {Altermagnetic lifting of {{Kramers}} spin degeneracy},\
  }\href {https://doi.org/10.1038/s41586-023-06907-7} {\bibfield  {journal}
  {\bibinfo  {journal} {Nature}\ }\textbf {\bibinfo {volume} {626}},\ \bibinfo
  {pages} {517} (\bibinfo {year} {2024})}\BibitemShut {NoStop}%
\bibitem [{\citenamefont {Reimers}\ \emph {et~al.}(2024)\citenamefont
  {Reimers}, \citenamefont {Odenbreit}, \citenamefont {{\v S}mejkal},
  \citenamefont {Strocov}, \citenamefont {Constantinou}, \citenamefont
  {Hellenes}, \citenamefont {Jaeschke~Ubiergo}, \citenamefont {Campos},
  \citenamefont {Bharadwaj}, \citenamefont {Chakraborty} \emph
  {et~al.}}]{reimers2024direct}%
  \BibitemOpen
  \bibfield  {author} {\bibinfo {author} {\bibfnamefont {S.}~\bibnamefont
  {Reimers}}, \bibinfo {author} {\bibfnamefont {L.}~\bibnamefont {Odenbreit}},
  \bibinfo {author} {\bibfnamefont {L.}~\bibnamefont {{\v S}mejkal}}, \bibinfo
  {author} {\bibfnamefont {V.~N.}\ \bibnamefont {Strocov}}, \bibinfo {author}
  {\bibfnamefont {P.}~\bibnamefont {Constantinou}}, \bibinfo {author}
  {\bibfnamefont {A.~B.}\ \bibnamefont {Hellenes}}, \bibinfo {author}
  {\bibfnamefont {R.}~\bibnamefont {Jaeschke~Ubiergo}}, \bibinfo {author}
  {\bibfnamefont {W.~H.}\ \bibnamefont {Campos}}, \bibinfo {author}
  {\bibfnamefont {V.~K.}\ \bibnamefont {Bharadwaj}}, \bibinfo {author}
  {\bibfnamefont {A.}~\bibnamefont {Chakraborty}}, \emph {et~al.},\ }\bibfield
  {title} {\bibinfo {title} {{Direct observation of altermagnetic band
  splitting in CrSb thin films}},\ }\href
  {https://doi.org/10.1038/s41467-024-46476-5} {\bibfield  {journal} {\bibinfo
  {journal} {Nat. Commun.}\ }\textbf {\bibinfo {volume} {15}},\ \bibinfo
  {pages} {2116} (\bibinfo {year} {2024})}\BibitemShut {NoStop}%
\bibitem [{\citenamefont {Tamang}\ \emph {et~al.}(2024)\citenamefont {Tamang},
  \citenamefont {Gurung}, \citenamefont {Rai}, \citenamefont {Brahimi},\ and\
  \citenamefont {Lounis}}]{tamang2024newly}%
  \BibitemOpen
  \bibfield  {author} {\bibinfo {author} {\bibfnamefont {R.}~\bibnamefont
  {Tamang}}, \bibinfo {author} {\bibfnamefont {S.}~\bibnamefont {Gurung}},
  \bibinfo {author} {\bibfnamefont {D.}~\bibnamefont {Rai}}, \bibinfo {author}
  {\bibfnamefont {S.}~\bibnamefont {Brahimi}},\ and\ \bibinfo {author}
  {\bibfnamefont {S.}~\bibnamefont {Lounis}},\ }\bibfield  {title} {\bibinfo
  {title} {Newly discovered magnetic phase: A brief review on altermagnets},\
  }\href {https://doi.org/10.48550/arXiv.2412.05377} {\bibfield  {journal}
  {\bibinfo  {journal} {arXiv:2412.05377}\ } (\bibinfo {year}
  {2024})}\BibitemShut {NoStop}%
\bibitem [{\citenamefont {Liu}\ \emph {et~al.}(2024{\natexlab{a}})\citenamefont
  {Liu}, \citenamefont {Zhan}, \citenamefont {Li}, \citenamefont {Liu},
  \citenamefont {Cheng}, \citenamefont {Shi}, \citenamefont {Deng},
  \citenamefont {Zhang}, \citenamefont {Li}, \citenamefont {Ding} \emph
  {et~al.}}]{liu2024absence}%
  \BibitemOpen
  \bibfield  {author} {\bibinfo {author} {\bibfnamefont {J.}~\bibnamefont
  {Liu}}, \bibinfo {author} {\bibfnamefont {J.}~\bibnamefont {Zhan}}, \bibinfo
  {author} {\bibfnamefont {T.}~\bibnamefont {Li}}, \bibinfo {author}
  {\bibfnamefont {J.}~\bibnamefont {Liu}}, \bibinfo {author} {\bibfnamefont
  {S.}~\bibnamefont {Cheng}}, \bibinfo {author} {\bibfnamefont
  {Y.}~\bibnamefont {Shi}}, \bibinfo {author} {\bibfnamefont {L.}~\bibnamefont
  {Deng}}, \bibinfo {author} {\bibfnamefont {M.}~\bibnamefont {Zhang}},
  \bibinfo {author} {\bibfnamefont {C.}~\bibnamefont {Li}}, \bibinfo {author}
  {\bibfnamefont {J.}~\bibnamefont {Ding}}, \emph {et~al.},\ }\bibfield
  {title} {\bibinfo {title} {{Absence of altermagnetic spin splitting character
  in rutile oxide RuO$_2$}},\ }\href
  {https://doi.org/10.1103/PhysRevLett.133.176401} {\bibfield  {journal}
  {\bibinfo  {journal} {Phys. Rev. Lett.}\ }\textbf {\bibinfo {volume} {133}},\
  \bibinfo {pages} {176401} (\bibinfo {year} {2024}{\natexlab{a}})}\BibitemShut
  {NoStop}%
\bibitem [{\citenamefont {Wang}\ \emph {et~al.}(2022)\citenamefont {Wang},
  \citenamefont {Chen}, \citenamefont {Ding}, \citenamefont {Zhang},
  \citenamefont {Dong},\ and\ \citenamefont {Wang}}]{wang2022magneto}%
  \BibitemOpen
  \bibfield  {author} {\bibinfo {author} {\bibfnamefont {N.}~\bibnamefont
  {Wang}}, \bibinfo {author} {\bibfnamefont {J.}~\bibnamefont {Chen}}, \bibinfo
  {author} {\bibfnamefont {N.}~\bibnamefont {Ding}}, \bibinfo {author}
  {\bibfnamefont {H.}~\bibnamefont {Zhang}}, \bibinfo {author} {\bibfnamefont
  {S.}~\bibnamefont {Dong}},\ and\ \bibinfo {author} {\bibfnamefont {S.-S.}\
  \bibnamefont {Wang}},\ }\bibfield  {title} {\bibinfo {title}
  {{Magneto-optical Kerr effect and magnetoelasticity in a weakly ferromagnetic
  RuF$_4$ monolayer}},\ }\href {https://doi.org/10.1103/PhysRevB.106.064435}
  {\bibfield  {journal} {\bibinfo  {journal} {Phys. Rev. B}\ }\textbf {\bibinfo
  {volume} {106}},\ \bibinfo {pages} {064435} (\bibinfo {year}
  {2022})}\BibitemShut {NoStop}%
\bibitem [{\citenamefont {Ding}\ \emph {et~al.}(2024)\citenamefont {Ding},
  \citenamefont {Jiang}, \citenamefont {Chen}, \citenamefont {Tao},
  \citenamefont {Liu}, \citenamefont {Li}, \citenamefont {Liu}, \citenamefont
  {Sun}, \citenamefont {Cheng}, \citenamefont {Liu} \emph
  {et~al.}}]{ding2024large}%
  \BibitemOpen
  \bibfield  {author} {\bibinfo {author} {\bibfnamefont {J.}~\bibnamefont
  {Ding}}, \bibinfo {author} {\bibfnamefont {Z.}~\bibnamefont {Jiang}},
  \bibinfo {author} {\bibfnamefont {X.}~\bibnamefont {Chen}}, \bibinfo {author}
  {\bibfnamefont {Z.}~\bibnamefont {Tao}}, \bibinfo {author} {\bibfnamefont
  {Z.}~\bibnamefont {Liu}}, \bibinfo {author} {\bibfnamefont {T.}~\bibnamefont
  {Li}}, \bibinfo {author} {\bibfnamefont {J.}~\bibnamefont {Liu}}, \bibinfo
  {author} {\bibfnamefont {J.}~\bibnamefont {Sun}}, \bibinfo {author}
  {\bibfnamefont {J.}~\bibnamefont {Cheng}}, \bibinfo {author} {\bibfnamefont
  {J.}~\bibnamefont {Liu}}, \emph {et~al.},\ }\bibfield  {title} {\bibinfo
  {title} {{Large band splitting in g-wave altermagnet CrSb}},\ }\href
  {https://doi.org/10.1103/PhysRevLett.133.206401} {\bibfield  {journal}
  {\bibinfo  {journal} {Phys. Rev. Lett.}\ }\textbf {\bibinfo {volume} {133}},\
  \bibinfo {pages} {206401} (\bibinfo {year} {2024})}\BibitemShut {NoStop}%
\bibitem [{\citenamefont {Chen}\ \emph {et~al.}(2025)\citenamefont {Chen},
  \citenamefont {Wang}, \citenamefont {Qin}, \citenamefont {Meng},
  \citenamefont {Zhou}, \citenamefont {Wang}, \citenamefont {Liu},
  \citenamefont {Zhao}, \citenamefont {Duan}, \citenamefont {Zhang},
  \citenamefont {Liu}, \citenamefont {Shao}, \citenamefont {Jiang},\ and\
  \citenamefont {Liu}}]{chen2024altermagnetic}%
  \BibitemOpen
  \bibfield  {author} {\bibinfo {author} {\bibfnamefont {H.}~\bibnamefont
  {Chen}}, \bibinfo {author} {\bibfnamefont {Z.-A.}\ \bibnamefont {Wang}},
  \bibinfo {author} {\bibfnamefont {P.}~\bibnamefont {Qin}}, \bibinfo {author}
  {\bibfnamefont {Z.}~\bibnamefont {Meng}}, \bibinfo {author} {\bibfnamefont
  {X.}~\bibnamefont {Zhou}}, \bibinfo {author} {\bibfnamefont {X.}~\bibnamefont
  {Wang}}, \bibinfo {author} {\bibfnamefont {L.}~\bibnamefont {Liu}}, \bibinfo
  {author} {\bibfnamefont {G.}~\bibnamefont {Zhao}}, \bibinfo {author}
  {\bibfnamefont {Z.}~\bibnamefont {Duan}}, \bibinfo {author} {\bibfnamefont
  {T.}~\bibnamefont {Zhang}}, \bibinfo {author} {\bibfnamefont
  {J.}~\bibnamefont {Liu}}, \bibinfo {author} {\bibfnamefont {D.-F.}\
  \bibnamefont {Shao}}, \bibinfo {author} {\bibfnamefont {C.}~\bibnamefont
  {Jiang}},\ and\ \bibinfo {author} {\bibfnamefont {Z.}~\bibnamefont {Liu}},\
  }\bibfield  {title} {\bibinfo {title} {Spin-splitting magnetoresistance in
  altermagnetic ruo2 thin films},\ }\href
  {https://doi.org/https://doi.org/10.1002/adma.202507764} {\bibfield
  {journal} {\bibinfo  {journal} {Advanced Materials}\ }\textbf {\bibinfo
  {volume} {37}},\ \bibinfo {pages} {2507764} (\bibinfo {year}
  {2025})}\BibitemShut {NoStop}%
\bibitem [{\citenamefont {Zhou}\ \emph
  {et~al.}(2024{\natexlab{a}})\citenamefont {Zhou}, \citenamefont {Feng},
  \citenamefont {Zhang}, \citenamefont {{\v S}mejkal}, \citenamefont {Sinova},
  \citenamefont {Mokrousov},\ and\ \citenamefont {Yao}}]{zhou2024crystal1}%
  \BibitemOpen
  \bibfield  {author} {\bibinfo {author} {\bibfnamefont {X.}~\bibnamefont
  {Zhou}}, \bibinfo {author} {\bibfnamefont {W.}~\bibnamefont {Feng}}, \bibinfo
  {author} {\bibfnamefont {R.-W.}\ \bibnamefont {Zhang}}, \bibinfo {author}
  {\bibfnamefont {L.}~\bibnamefont {{\v S}mejkal}}, \bibinfo {author}
  {\bibfnamefont {J.}~\bibnamefont {Sinova}}, \bibinfo {author} {\bibfnamefont
  {Y.}~\bibnamefont {Mokrousov}},\ and\ \bibinfo {author} {\bibfnamefont
  {Y.}~\bibnamefont {Yao}},\ }\bibfield  {title} {\bibinfo {title} {{Crystal
  {{Thermal Transport}} in {{Altermagnetic}} RuO$_2$}},\ }\href
  {https://doi.org/10.1103/PhysRevLett.132.056701} {\bibfield  {journal}
  {\bibinfo  {journal} {Phys. Rev. Lett.}\ }\textbf {\bibinfo {volume} {132}},\
  \bibinfo {pages} {056701} (\bibinfo {year} {2024}{\natexlab{a}})}\BibitemShut
  {NoStop}%
\bibitem [{\citenamefont {Attias}\ \emph {et~al.}(2024)\citenamefont {Attias},
  \citenamefont {Levchenko},\ and\ \citenamefont
  {Khodas}}]{attias2024intrinsic}%
  \BibitemOpen
  \bibfield  {author} {\bibinfo {author} {\bibfnamefont {L.}~\bibnamefont
  {Attias}}, \bibinfo {author} {\bibfnamefont {A.}~\bibnamefont {Levchenko}},\
  and\ \bibinfo {author} {\bibfnamefont {M.}~\bibnamefont {Khodas}},\
  }\bibfield  {title} {\bibinfo {title} {Intrinsic anomalous {{Hall}} effect in
  altermagnets},\ }\href {https://doi.org/10.1103/PhysRevB.110.094425}
  {\bibfield  {journal} {\bibinfo  {journal} {Phys. Rev. B}\ }\textbf {\bibinfo
  {volume} {110}},\ \bibinfo {pages} {094425} (\bibinfo {year}
  {2024})}\BibitemShut {NoStop}%
\bibitem [{\citenamefont {Han}\ \emph {et~al.}(2024)\citenamefont {Han},
  \citenamefont {Fu}, \citenamefont {He}, \citenamefont {Zhu}, \citenamefont
  {Dai}, \citenamefont {Yang}, \citenamefont {Zhu}, \citenamefont {Bai},
  \citenamefont {Chen}, \citenamefont {Wan} \emph
  {et~al.}}]{han2024observation}%
  \BibitemOpen
  \bibfield  {author} {\bibinfo {author} {\bibfnamefont {L.}~\bibnamefont
  {Han}}, \bibinfo {author} {\bibfnamefont {X.}~\bibnamefont {Fu}}, \bibinfo
  {author} {\bibfnamefont {W.}~\bibnamefont {He}}, \bibinfo {author}
  {\bibfnamefont {Y.}~\bibnamefont {Zhu}}, \bibinfo {author} {\bibfnamefont
  {J.}~\bibnamefont {Dai}}, \bibinfo {author} {\bibfnamefont {W.}~\bibnamefont
  {Yang}}, \bibinfo {author} {\bibfnamefont {W.}~\bibnamefont {Zhu}}, \bibinfo
  {author} {\bibfnamefont {H.}~\bibnamefont {Bai}}, \bibinfo {author}
  {\bibfnamefont {C.}~\bibnamefont {Chen}}, \bibinfo {author} {\bibfnamefont
  {C.}~\bibnamefont {Wan}}, \emph {et~al.},\ }\bibfield  {title} {\bibinfo
  {title} {{Observation of non-volatile anomalous Nernst effect in altermagnet
  with collinear N\'eel vector}},\ }\href
  {https://doi.org/10.48550/arXiv.2403.13427} {\bibfield  {journal} {\bibinfo
  {journal} {arXiv:2403.13427}\ } (\bibinfo {year} {2024})}\BibitemShut
  {NoStop}%
\bibitem [{\citenamefont {Zeng}\ and\ \citenamefont
  {Zhao}(2024)}]{zeng2024description}%
  \BibitemOpen
  \bibfield  {author} {\bibinfo {author} {\bibfnamefont {S.}~\bibnamefont
  {Zeng}}\ and\ \bibinfo {author} {\bibfnamefont {Y.-J.}\ \bibnamefont
  {Zhao}},\ }\bibfield  {title} {\bibinfo {title} {Description of
  two-dimensional altermagnetism: Categorization using spin group theory},\
  }\href {https://doi.org/10.1103/PhysRevB.110.054406} {\bibfield  {journal}
  {\bibinfo  {journal} {Phys. Rev. B}\ }\textbf {\bibinfo {volume} {110}},\
  \bibinfo {pages} {054406} (\bibinfo {year} {2024})}\BibitemShut {NoStop}%
\bibitem [{\citenamefont {Xiao}\ \emph {et~al.}(2024)\citenamefont {Xiao},
  \citenamefont {Zhao}, \citenamefont {Li}, \citenamefont {Shindou},\ and\
  \citenamefont {Song}}]{xiao2024spin}%
  \BibitemOpen
  \bibfield  {author} {\bibinfo {author} {\bibfnamefont {Z.}~\bibnamefont
  {Xiao}}, \bibinfo {author} {\bibfnamefont {J.}~\bibnamefont {Zhao}}, \bibinfo
  {author} {\bibfnamefont {Y.}~\bibnamefont {Li}}, \bibinfo {author}
  {\bibfnamefont {R.}~\bibnamefont {Shindou}},\ and\ \bibinfo {author}
  {\bibfnamefont {Z.-D.}\ \bibnamefont {Song}},\ }\bibfield  {title} {\bibinfo
  {title} {Spin space groups: Full classification and applications},\ }\href
  {https://doi.org/10.1103/PhysRevX.14.031037} {\bibfield  {journal} {\bibinfo
  {journal} {Phys. Rev. X}\ }\textbf {\bibinfo {volume} {14}},\ \bibinfo
  {pages} {031037} (\bibinfo {year} {2024})}\BibitemShut {NoStop}%
\bibitem [{\citenamefont {Jiang}\ \emph {et~al.}(2024)\citenamefont {Jiang},
  \citenamefont {Song}, \citenamefont {Zhu}, \citenamefont {Fang},
  \citenamefont {Weng}, \citenamefont {Liu}, \citenamefont {Yang},\ and\
  \citenamefont {Fang}}]{jiang2024enumeration}%
  \BibitemOpen
  \bibfield  {author} {\bibinfo {author} {\bibfnamefont {Y.}~\bibnamefont
  {Jiang}}, \bibinfo {author} {\bibfnamefont {Z.}~\bibnamefont {Song}},
  \bibinfo {author} {\bibfnamefont {T.}~\bibnamefont {Zhu}}, \bibinfo {author}
  {\bibfnamefont {Z.}~\bibnamefont {Fang}}, \bibinfo {author} {\bibfnamefont
  {H.}~\bibnamefont {Weng}}, \bibinfo {author} {\bibfnamefont {Z.-X.}\
  \bibnamefont {Liu}}, \bibinfo {author} {\bibfnamefont {J.}~\bibnamefont
  {Yang}},\ and\ \bibinfo {author} {\bibfnamefont {C.}~\bibnamefont {Fang}},\
  }\bibfield  {title} {\bibinfo {title} {Enumeration of spin-space groups:
  Toward a complete description of symmetries of magnetic orders},\ }\href
  {https://doi.org/10.1103/PhysRevX.14.031039} {\bibfield  {journal} {\bibinfo
  {journal} {Phys. Rev. X}\ }\textbf {\bibinfo {volume} {14}},\ \bibinfo
  {pages} {031039} (\bibinfo {year} {2024})}\BibitemShut {NoStop}%
\bibitem [{\citenamefont {Xiao}\ \emph {et~al.}(2025)\citenamefont {Xiao},
  \citenamefont {Li}, \citenamefont {Han}, \citenamefont {Gan}, \citenamefont
  {Yang}, \citenamefont {Shao}, \citenamefont {Zhang}, \citenamefont {Gao},
  \citenamefont {Tian},\ and\ \citenamefont {Zhou}}]{xiao2024anomalous}%
  \BibitemOpen
  \bibfield  {author} {\bibinfo {author} {\bibfnamefont {R.-C.}\ \bibnamefont
  {Xiao}}, \bibinfo {author} {\bibfnamefont {H.}~\bibnamefont {Li}}, \bibinfo
  {author} {\bibfnamefont {H.}~\bibnamefont {Han}}, \bibinfo {author}
  {\bibfnamefont {W.}~\bibnamefont {Gan}}, \bibinfo {author} {\bibfnamefont
  {M.}~\bibnamefont {Yang}}, \bibinfo {author} {\bibfnamefont {D.-F.}\
  \bibnamefont {Shao}}, \bibinfo {author} {\bibfnamefont {S.-H.}\ \bibnamefont
  {Zhang}}, \bibinfo {author} {\bibfnamefont {Y.}~\bibnamefont {Gao}}, \bibinfo
  {author} {\bibfnamefont {M.}~\bibnamefont {Tian}},\ and\ \bibinfo {author}
  {\bibfnamefont {J.}~\bibnamefont {Zhou}},\ }\bibfield  {title} {\bibinfo
  {title} {Anomalous-hall néel textures in altermagnetic materials},\ }\href
  {https://doi.org/10.1007/s11433-025-2769-6} {\bibfield  {journal} {\bibinfo
  {journal} {Science China Physics, Mechanics \& Astronomy}\ }\textbf {\bibinfo
  {volume} {69}},\ \bibinfo {pages} {217511} (\bibinfo {year}
  {2025})}\BibitemShut {NoStop}%
\bibitem [{\citenamefont {Leeb}\ \emph {et~al.}(2024)\citenamefont {Leeb},
  \citenamefont {Mook}, \citenamefont {{\v{S}}mejkal},\ and\ \citenamefont
  {Knolle}}]{leeb2024spontaneous}%
  \BibitemOpen
  \bibfield  {author} {\bibinfo {author} {\bibfnamefont {V.}~\bibnamefont
  {Leeb}}, \bibinfo {author} {\bibfnamefont {A.}~\bibnamefont {Mook}}, \bibinfo
  {author} {\bibfnamefont {L.}~\bibnamefont {{\v{S}}mejkal}},\ and\ \bibinfo
  {author} {\bibfnamefont {J.}~\bibnamefont {Knolle}},\ }\bibfield  {title}
  {\bibinfo {title} {Spontaneous formation of altermagnetism from orbital
  ordering},\ }\href {https://doi.org/10.1103/PhysRevLett.132.236701}
  {\bibfield  {journal} {\bibinfo  {journal} {Phys. Rev. Lett.}\ }\textbf
  {\bibinfo {volume} {132}},\ \bibinfo {pages} {236701} (\bibinfo {year}
  {2024})}\BibitemShut {NoStop}%
\bibitem [{\citenamefont {Zhang}\ \emph
  {et~al.}(2025{\natexlab{a}})\citenamefont {Zhang}, \citenamefont {Cheng},
  \citenamefont {Yin}, \citenamefont {Liu}, \citenamefont {Deng}, \citenamefont
  {Qiao}, \citenamefont {Shi}, \citenamefont {Zhang}, \citenamefont {Lin},
  \citenamefont {Liu}, \citenamefont {Ye}, \citenamefont {Huang}, \citenamefont
  {Meng}, \citenamefont {Zhang}, \citenamefont {Okuda}, \citenamefont
  {Shimada}, \citenamefont {Cui}, \citenamefont {Zhao}, \citenamefont {Cao},
  \citenamefont {Qiao}, \citenamefont {Liu},\ and\ \citenamefont
  {Chen}}]{zhang2025Crystal}%
  \BibitemOpen
  \bibfield  {author} {\bibinfo {author} {\bibfnamefont {F.}~\bibnamefont
  {Zhang}}, \bibinfo {author} {\bibfnamefont {X.}~\bibnamefont {Cheng}},
  \bibinfo {author} {\bibfnamefont {Z.}~\bibnamefont {Yin}}, \bibinfo {author}
  {\bibfnamefont {C.}~\bibnamefont {Liu}}, \bibinfo {author} {\bibfnamefont
  {L.}~\bibnamefont {Deng}}, \bibinfo {author} {\bibfnamefont {Y.}~\bibnamefont
  {Qiao}}, \bibinfo {author} {\bibfnamefont {Z.}~\bibnamefont {Shi}}, \bibinfo
  {author} {\bibfnamefont {S.}~\bibnamefont {Zhang}}, \bibinfo {author}
  {\bibfnamefont {J.}~\bibnamefont {Lin}}, \bibinfo {author} {\bibfnamefont
  {Z.}~\bibnamefont {Liu}}, \bibinfo {author} {\bibfnamefont {M.}~\bibnamefont
  {Ye}}, \bibinfo {author} {\bibfnamefont {Y.}~\bibnamefont {Huang}}, \bibinfo
  {author} {\bibfnamefont {X.}~\bibnamefont {Meng}}, \bibinfo {author}
  {\bibfnamefont {C.}~\bibnamefont {Zhang}}, \bibinfo {author} {\bibfnamefont
  {T.}~\bibnamefont {Okuda}}, \bibinfo {author} {\bibfnamefont
  {K.}~\bibnamefont {Shimada}}, \bibinfo {author} {\bibfnamefont
  {S.}~\bibnamefont {Cui}}, \bibinfo {author} {\bibfnamefont {Y.}~\bibnamefont
  {Zhao}}, \bibinfo {author} {\bibfnamefont {G.-H.}\ \bibnamefont {Cao}},
  \bibinfo {author} {\bibfnamefont {S.}~\bibnamefont {Qiao}}, \bibinfo {author}
  {\bibfnamefont {J.}~\bibnamefont {Liu}},\ and\ \bibinfo {author}
  {\bibfnamefont {C.}~\bibnamefont {Chen}},\ }\bibfield  {title} {\bibinfo
  {title} {Crystal-symmetry-paired spin--valley locking in a layered
  room-temperature metallic altermagnet candidate},\ }\href
  {https://doi.org/10.1038/s41567-025-02864-2} {\bibfield  {journal} {\bibinfo
  {journal} {Nat. Phys.}\ }\textbf {\bibinfo {volume} {21}},\ \bibinfo {pages}
  {760} (\bibinfo {year} {2025}{\natexlab{a}})}\BibitemShut {NoStop}%
\bibitem [{\citenamefont {Qian}\ \emph {et~al.}(2025)\citenamefont {Qian},
  \citenamefont {Yang}, \citenamefont {Liu},\ and\ \citenamefont
  {Wu}}]{qian2025fragile}%
  \BibitemOpen
  \bibfield  {author} {\bibinfo {author} {\bibfnamefont {Z.}~\bibnamefont
  {Qian}}, \bibinfo {author} {\bibfnamefont {Y.}~\bibnamefont {Yang}}, \bibinfo
  {author} {\bibfnamefont {S.}~\bibnamefont {Liu}},\ and\ \bibinfo {author}
  {\bibfnamefont {C.}~\bibnamefont {Wu}},\ }\bibfield  {title} {\bibinfo
  {title} {Fragile unconventional magnetism in ${\mathrm{ruo}}_{2}$ by
  proximity to landau-pomeranchuk instability},\ }\href
  {https://doi.org/10.1103/PhysRevB.111.174425} {\bibfield  {journal} {\bibinfo
   {journal} {Phys. Rev. B}\ }\textbf {\bibinfo {volume} {111}},\ \bibinfo
  {pages} {174425} (\bibinfo {year} {2025})}\BibitemShut {NoStop}%
\bibitem [{\citenamefont {Liu}\ \emph {et~al.}(2025)\citenamefont {Liu},
  \citenamefont {Dai},\ and\ \citenamefont {Bl{\"u}gel}}]{liu2025different}%
  \BibitemOpen
  \bibfield  {author} {\bibinfo {author} {\bibfnamefont {Q.}~\bibnamefont
  {Liu}}, \bibinfo {author} {\bibfnamefont {X.}~\bibnamefont {Dai}},\ and\
  \bibinfo {author} {\bibfnamefont {S.}~\bibnamefont {Bl{\"u}gel}},\ }\bibfield
   {title} {\bibinfo {title} {Different facets of unconventional magnetism},\
  }\href {https://doi.org/10.1038/s41567-024-02750-3} {\bibfield  {journal}
  {\bibinfo  {journal} {Nat. Phys.}\ }\textbf {\bibinfo {volume} {21}},\
  \bibinfo {pages} {329} (\bibinfo {year} {2025})}\BibitemShut {NoStop}%
\bibitem [{\citenamefont {Liu}\ and\ \citenamefont
  {Medhekar}(2025)}]{liu2025d-wave}%
  \BibitemOpen
  \bibfield  {author} {\bibinfo {author} {\bibfnamefont {Z.}~\bibnamefont
  {Liu}}\ and\ \bibinfo {author} {\bibfnamefont {N.~V.}\ \bibnamefont
  {Medhekar}},\ }\bibfield  {title} {\bibinfo {title} {d-wave polarization-spin
  locking in two-dimensional altermagnets},\ }\href
  {https://doi.org/10.1021/acs.nanolett.5c01178} {\bibfield  {journal}
  {\bibinfo  {journal} {Nano Letters}\ }\textbf {\bibinfo {volume} {25}},\
  \bibinfo {pages} {13411} (\bibinfo {year} {2025})}\BibitemShut {NoStop}%
\bibitem [{\citenamefont {Wang}\ \emph
  {et~al.}(2025{\natexlab{a}})\citenamefont {Wang}, \citenamefont {Li},
  \citenamefont {Zhang}, \citenamefont {Lu}, \citenamefont {Tian},
  \citenamefont {Sun}, \citenamefont {Tsymbal}, \citenamefont {Wang},
  \citenamefont {Du},\ and\ \citenamefont {Shao}}]{wang2025giant}%
  \BibitemOpen
  \bibfield  {author} {\bibinfo {author} {\bibfnamefont {Z.-A.}\ \bibnamefont
  {Wang}}, \bibinfo {author} {\bibfnamefont {B.}~\bibnamefont {Li}}, \bibinfo
  {author} {\bibfnamefont {S.-S.}\ \bibnamefont {Zhang}}, \bibinfo {author}
  {\bibfnamefont {W.-J.}\ \bibnamefont {Lu}}, \bibinfo {author} {\bibfnamefont
  {M.}~\bibnamefont {Tian}}, \bibinfo {author} {\bibfnamefont {Y.-P.}\
  \bibnamefont {Sun}}, \bibinfo {author} {\bibfnamefont {E.~Y.}\ \bibnamefont
  {Tsymbal}}, \bibinfo {author} {\bibfnamefont {K.}~\bibnamefont {Wang}},
  \bibinfo {author} {\bibfnamefont {H.}~\bibnamefont {Du}},\ and\ \bibinfo
  {author} {\bibfnamefont {D.-F.}\ \bibnamefont {Shao}},\ }\bibfield  {title}
  {\bibinfo {title} {Giant uncompensated magnon spin currents in x-type
  magnets},\ }\href {https://doi.org/10.48550/arXiv.2502.13511} {\bibfield
  {journal} {\bibinfo  {journal} {arXiv:2502.13511}\ } (\bibinfo {year}
  {2025}{\natexlab{a}})}\BibitemShut {NoStop}%
\bibitem [{\citenamefont {Gao}\ \emph {et~al.}(2025)\citenamefont {Gao},
  \citenamefont {Qu}, \citenamefont {Zeng}, \citenamefont {Liu}, \citenamefont
  {Wen}, \citenamefont {Sun}, \citenamefont {Guo},\ and\ \citenamefont
  {Lu}}]{gao2025ai}%
  \BibitemOpen
  \bibfield  {author} {\bibinfo {author} {\bibfnamefont {Z.-F.}\ \bibnamefont
  {Gao}}, \bibinfo {author} {\bibfnamefont {S.}~\bibnamefont {Qu}}, \bibinfo
  {author} {\bibfnamefont {B.}~\bibnamefont {Zeng}}, \bibinfo {author}
  {\bibfnamefont {Y.}~\bibnamefont {Liu}}, \bibinfo {author} {\bibfnamefont
  {J.-R.}\ \bibnamefont {Wen}}, \bibinfo {author} {\bibfnamefont
  {H.}~\bibnamefont {Sun}}, \bibinfo {author} {\bibfnamefont {P.-J.}\
  \bibnamefont {Guo}},\ and\ \bibinfo {author} {\bibfnamefont {Z.-Y.}\
  \bibnamefont {Lu}},\ }\bibfield  {title} {\bibinfo {title} {{AI-accelerated
  discovery of altermagnetic materials}},\ }\href
  {https://doi.org/10.1093/nsr/nwaf066} {\bibfield  {journal} {\bibinfo
  {journal} {Nat. Sci. Rev.}\ }\textbf {\bibinfo {volume} {12}},\ \bibinfo
  {pages} {nwaf066} (\bibinfo {year} {2025})}\BibitemShut {NoStop}%
\bibitem [{\citenamefont {Yuan}\ \emph {et~al.}(2025)\citenamefont {Yuan},
  \citenamefont {Pan},\ and\ \citenamefont {Wu}}]{yuan2025unconventional}%
  \BibitemOpen
  \bibfield  {author} {\bibinfo {author} {\bibfnamefont {J.-K.}\ \bibnamefont
  {Yuan}}, \bibinfo {author} {\bibfnamefont {Z.}~\bibnamefont {Pan}},\ and\
  \bibinfo {author} {\bibfnamefont {C.}~\bibnamefont {Wu}},\ }\bibfield
  {title} {\bibinfo {title} {Unconventional magnetism in spin-orbit coupled
  systems},\ }\href {https://doi.org/10.48550/arXiv.2504.14577} {\bibfield
  {journal} {\bibinfo  {journal} {arXiv:2504.14577}\ } (\bibinfo {year}
  {2025})}\BibitemShut {NoStop}%
\bibitem [{\citenamefont {Hu}\ \emph {et~al.}(2025)\citenamefont {Hu},
  \citenamefont {Cheng}, \citenamefont {Huang},\ and\ \citenamefont
  {Liu}}]{hu2025Catalog}%
  \BibitemOpen
  \bibfield  {author} {\bibinfo {author} {\bibfnamefont {M.}~\bibnamefont
  {Hu}}, \bibinfo {author} {\bibfnamefont {X.}~\bibnamefont {Cheng}}, \bibinfo
  {author} {\bibfnamefont {Z.}~\bibnamefont {Huang}},\ and\ \bibinfo {author}
  {\bibfnamefont {J.}~\bibnamefont {Liu}},\ }\bibfield  {title} {\bibinfo
  {title} {Catalog of {{C}} -{{Paired Spin-Momentum Locking}} in
  {{Antiferromagnetic Systems}}},\ }\href
  {https://doi.org/10.1103/PhysRevX.15.021083} {\bibfield  {journal} {\bibinfo
  {journal} {Phys. Rev. X}\ }\textbf {\bibinfo {volume} {15}},\ \bibinfo
  {pages} {021083} (\bibinfo {year} {2025})}\BibitemShut {NoStop}%
\bibitem [{\citenamefont {Hayami}\ \emph {et~al.}(2020)\citenamefont {Hayami},
  \citenamefont {Yanagi},\ and\ \citenamefont {Kusunose}}]{hayami2020bottom}%
  \BibitemOpen
  \bibfield  {author} {\bibinfo {author} {\bibfnamefont {S.}~\bibnamefont
  {Hayami}}, \bibinfo {author} {\bibfnamefont {Y.}~\bibnamefont {Yanagi}},\
  and\ \bibinfo {author} {\bibfnamefont {H.}~\bibnamefont {Kusunose}},\
  }\bibfield  {title} {\bibinfo {title} {{Bottom-up design of spin-split and
  reshaped electronic band structures in antiferromagnets without spin-orbit
  coupling: Procedure on the basis of augmented multipoles}},\ }\href
  {https://doi.org/10.1103/PhysRevB.102.144441} {\bibfield  {journal} {\bibinfo
   {journal} {Phys. Rev. B}\ }\textbf {\bibinfo {volume} {102}},\ \bibinfo
  {pages} {144441} (\bibinfo {year} {2020})}\BibitemShut {NoStop}%
\bibitem [{\citenamefont {Bai}\ \emph {et~al.}(2024)\citenamefont {Bai},
  \citenamefont {Feng}, \citenamefont {Liu}, \citenamefont {{\v S}mejkal},
  \citenamefont {Mokrousov},\ and\ \citenamefont
  {Yao}}]{bai2024altermagnetism}%
  \BibitemOpen
  \bibfield  {author} {\bibinfo {author} {\bibfnamefont {L.}~\bibnamefont
  {Bai}}, \bibinfo {author} {\bibfnamefont {W.}~\bibnamefont {Feng}}, \bibinfo
  {author} {\bibfnamefont {S.}~\bibnamefont {Liu}}, \bibinfo {author}
  {\bibfnamefont {L.}~\bibnamefont {{\v S}mejkal}}, \bibinfo {author}
  {\bibfnamefont {Y.}~\bibnamefont {Mokrousov}},\ and\ \bibinfo {author}
  {\bibfnamefont {Y.}~\bibnamefont {Yao}},\ }\bibfield  {title} {\bibinfo
  {title} {Altermagnetism: {{Exploring New Frontiers}} in {{Magnetism}} and
  {{Spintronics}}},\ }\href {https://doi.org/10.1002/adfm.202409327} {\bibfield
   {journal} {\bibinfo  {journal} {Adv. Funct. Mater.}\ }\textbf {\bibinfo
  {volume} {34}},\ \bibinfo {pages} {2409327} (\bibinfo {year}
  {2024})}\BibitemShut {NoStop}%
\bibitem [{\citenamefont {Song}\ \emph {et~al.}(2025)\citenamefont {Song},
  \citenamefont {Bai}, \citenamefont {Zhou}, \citenamefont {Han}, \citenamefont
  {Reichlova}, \citenamefont {Dil}, \citenamefont {Liu}, \citenamefont {Chen},\
  and\ \citenamefont {Pan}}]{song2025altermagnets}%
  \BibitemOpen
  \bibfield  {author} {\bibinfo {author} {\bibfnamefont {C.}~\bibnamefont
  {Song}}, \bibinfo {author} {\bibfnamefont {H.}~\bibnamefont {Bai}}, \bibinfo
  {author} {\bibfnamefont {Z.}~\bibnamefont {Zhou}}, \bibinfo {author}
  {\bibfnamefont {L.}~\bibnamefont {Han}}, \bibinfo {author} {\bibfnamefont
  {H.}~\bibnamefont {Reichlova}}, \bibinfo {author} {\bibfnamefont {J.~H.}\
  \bibnamefont {Dil}}, \bibinfo {author} {\bibfnamefont {J.}~\bibnamefont
  {Liu}}, \bibinfo {author} {\bibfnamefont {X.}~\bibnamefont {Chen}},\ and\
  \bibinfo {author} {\bibfnamefont {F.}~\bibnamefont {Pan}},\ }\bibfield
  {title} {\bibinfo {title} {Altermagnets as a new class of functional
  materials},\ }\href {https://doi.org/10.1038/s41578-025-00779-1} {\bibfield
  {journal} {\bibinfo  {journal} {Nat. Rev. Mater.}\ } (\bibinfo {year}
  {2025})}\BibitemShut {NoStop}%
\bibitem [{\citenamefont {Fender}\ \emph {et~al.}(2025)\citenamefont {Fender},
  \citenamefont {Gonzalez},\ and\ \citenamefont
  {Bediako}}]{fender2025altermagnetism}%
  \BibitemOpen
  \bibfield  {author} {\bibinfo {author} {\bibfnamefont {S.~S.}\ \bibnamefont
  {Fender}}, \bibinfo {author} {\bibfnamefont {O.}~\bibnamefont {Gonzalez}},\
  and\ \bibinfo {author} {\bibfnamefont {D.~K.}\ \bibnamefont {Bediako}},\
  }\bibfield  {title} {\bibinfo {title} {Altermagnetism: {{A Chemical
  Perspective}}},\ }\href {https://doi.org/10.1021/jacs.4c14503} {\bibfield
  {journal} {\bibinfo  {journal} {J. Am. Chem. Soc.}\ }\textbf {\bibinfo
  {volume} {147}},\ \bibinfo {pages} {2257} (\bibinfo {year}
  {2025})}\BibitemShut {NoStop}%
\bibitem [{\citenamefont {Peng}\ \emph {et~al.}(2025)\citenamefont {Peng},
  \citenamefont {Yang}, \citenamefont {Hu}, \citenamefont {Ong}, \citenamefont
  {Ho}, \citenamefont {Lau}, \citenamefont {Liu},\ and\ \citenamefont
  {Ang}}]{peng2025all}%
  \BibitemOpen
  \bibfield  {author} {\bibinfo {author} {\bibfnamefont {R.}~\bibnamefont
  {Peng}}, \bibinfo {author} {\bibfnamefont {J.}~\bibnamefont {Yang}}, \bibinfo
  {author} {\bibfnamefont {L.}~\bibnamefont {Hu}}, \bibinfo {author}
  {\bibfnamefont {W.-L.}\ \bibnamefont {Ong}}, \bibinfo {author} {\bibfnamefont
  {P.}~\bibnamefont {Ho}}, \bibinfo {author} {\bibfnamefont {C.~S.}\
  \bibnamefont {Lau}}, \bibinfo {author} {\bibfnamefont {J.}~\bibnamefont
  {Liu}},\ and\ \bibinfo {author} {\bibfnamefont {Y.~S.}\ \bibnamefont {Ang}},\
  }\bibfield  {title} {\bibinfo {title} {All-electrical layer-spintronics in
  altermagnetic bilayers},\ }\href {https://doi.org/10.1093/nsr/nwaf066}
  {\bibfield  {journal} {\bibinfo  {journal} {Mater. Horiz.}\ }\textbf
  {\bibinfo {volume} {12}},\ \bibinfo {pages} {2197} (\bibinfo {year}
  {2025})}\BibitemShut {NoStop}%
\bibitem [{\citenamefont {Santhosh}\ \emph {et~al.}(2025)\citenamefont
  {Santhosh}, \citenamefont {Corbae}, \citenamefont {Yánez-Parreño},
  \citenamefont {Ghosh}, \citenamefont {Jensen}, \citenamefont {Fedorov},
  \citenamefont {Hashimoto}, \citenamefont {Lu}, \citenamefont {Borchers},
  \citenamefont {Grutter}, \citenamefont {Charlton}, \citenamefont {Islam},
  \citenamefont {Golovanova}, \citenamefont {Zhao}, \citenamefont {Tauraso},
  \citenamefont {Richardella}, \citenamefont {Yan}, \citenamefont {Mkhoyan},
  \citenamefont {Palmstrøm}, \citenamefont {Ou},\ and\ \citenamefont
  {Samarth}}]{santhosh2025altermagnetic}%
  \BibitemOpen
  \bibfield  {author} {\bibinfo {author} {\bibfnamefont {S.}~\bibnamefont
  {Santhosh}}, \bibinfo {author} {\bibfnamefont {P.}~\bibnamefont {Corbae}},
  \bibinfo {author} {\bibfnamefont {W.~J.}\ \bibnamefont {Yánez-Parreño}},
  \bibinfo {author} {\bibfnamefont {S.}~\bibnamefont {Ghosh}}, \bibinfo
  {author} {\bibfnamefont {C.~J.}\ \bibnamefont {Jensen}}, \bibinfo {author}
  {\bibfnamefont {A.~V.}\ \bibnamefont {Fedorov}}, \bibinfo {author}
  {\bibfnamefont {M.}~\bibnamefont {Hashimoto}}, \bibinfo {author}
  {\bibfnamefont {D.}~\bibnamefont {Lu}}, \bibinfo {author} {\bibfnamefont
  {J.~A.}\ \bibnamefont {Borchers}}, \bibinfo {author} {\bibfnamefont {A.~J.}\
  \bibnamefont {Grutter}}, \bibinfo {author} {\bibfnamefont {T.~R.}\
  \bibnamefont {Charlton}}, \bibinfo {author} {\bibfnamefont {S.}~\bibnamefont
  {Islam}}, \bibinfo {author} {\bibfnamefont {D.}~\bibnamefont {Golovanova}},
  \bibinfo {author} {\bibfnamefont {Y.}~\bibnamefont {Zhao}}, \bibinfo {author}
  {\bibfnamefont {A.}~\bibnamefont {Tauraso}}, \bibinfo {author} {\bibfnamefont
  {A.}~\bibnamefont {Richardella}}, \bibinfo {author} {\bibfnamefont
  {B.}~\bibnamefont {Yan}}, \bibinfo {author} {\bibfnamefont {K.~A.}\
  \bibnamefont {Mkhoyan}}, \bibinfo {author} {\bibfnamefont {C.~J.}\
  \bibnamefont {Palmstrøm}}, \bibinfo {author} {\bibfnamefont
  {Y.}~\bibnamefont {Ou}},\ and\ \bibinfo {author} {\bibfnamefont
  {N.}~\bibnamefont {Samarth}},\ }\bibfield  {title} {\bibinfo {title}
  {Altermagnetic band splitting in 10 nm epitaxial crsb thin films},\ }\href
  {https://doi.org/https://doi.org/10.1002/adma.202508977} {\bibfield
  {journal} {\bibinfo  {journal} {Advanced Materials}\ }\textbf {\bibinfo
  {volume} {37}},\ \bibinfo {pages} {e08977} (\bibinfo {year} {2025})},\
  \Eprint
  {https://arxiv.org/abs/https://advanced.onlinelibrary.wiley.com/doi/pdf/10.1002/adma.202508977}
  {https://advanced.onlinelibrary.wiley.com/doi/pdf/10.1002/adma.202508977}
  \BibitemShut {NoStop}%
\bibitem [{\citenamefont {Samanta}\ \emph {et~al.}(2020)\citenamefont
  {Samanta}, \citenamefont {Ležaić}, \citenamefont {Merte}, \citenamefont
  {Freimuth}, \citenamefont {Blügel},\ and\ \citenamefont
  {Mokrousov}}]{samanta2020crystal}%
  \BibitemOpen
  \bibfield  {author} {\bibinfo {author} {\bibfnamefont {K.}~\bibnamefont
  {Samanta}}, \bibinfo {author} {\bibfnamefont {M.}~\bibnamefont {Ležaić}},
  \bibinfo {author} {\bibfnamefont {M.}~\bibnamefont {Merte}}, \bibinfo
  {author} {\bibfnamefont {F.}~\bibnamefont {Freimuth}}, \bibinfo {author}
  {\bibfnamefont {S.}~\bibnamefont {Blügel}},\ and\ \bibinfo {author}
  {\bibfnamefont {Y.}~\bibnamefont {Mokrousov}},\ }\bibfield  {title} {\bibinfo
  {title} {{Crystal Hall and crystal magneto-optical effect in thin films of
  SrRuO$_3$}},\ }\href {https://doi.org/10.1063/5.0005017} {\bibfield
  {journal} {\bibinfo  {journal} {J. Appl. Phys.}\ }\textbf {\bibinfo {volume}
  {127}},\ \bibinfo {pages} {213904} (\bibinfo {year} {2020})}\BibitemShut
  {NoStop}%
\bibitem [{\citenamefont {Zhang}\ \emph
  {et~al.}(2025{\natexlab{b}})\citenamefont {Zhang}, \citenamefont {Bai},
  \citenamefont {Zhang}, \citenamefont {Chen}, \citenamefont {Han},
  \citenamefont {Liang}, \citenamefont {Chu}, \citenamefont {Dai},
  \citenamefont {Sawicki}, \citenamefont {Pan},\ and\ \citenamefont
  {Song}}]{zhang2024probing}%
  \BibitemOpen
  \bibfield  {author} {\bibinfo {author} {\bibfnamefont {Y.-C.}\ \bibnamefont
  {Zhang}}, \bibinfo {author} {\bibfnamefont {H.}~\bibnamefont {Bai}}, \bibinfo
  {author} {\bibfnamefont {D.-H.}\ \bibnamefont {Zhang}}, \bibinfo {author}
  {\bibfnamefont {C.}~\bibnamefont {Chen}}, \bibinfo {author} {\bibfnamefont
  {L.}~\bibnamefont {Han}}, \bibinfo {author} {\bibfnamefont {S.-X.}\
  \bibnamefont {Liang}}, \bibinfo {author} {\bibfnamefont {R.-Y.}\ \bibnamefont
  {Chu}}, \bibinfo {author} {\bibfnamefont {J.-K.}\ \bibnamefont {Dai}},
  \bibinfo {author} {\bibfnamefont {M.}~\bibnamefont {Sawicki}}, \bibinfo
  {author} {\bibfnamefont {F.}~\bibnamefont {Pan}},\ and\ \bibinfo {author}
  {\bibfnamefont {C.}~\bibnamefont {Song}},\ }\bibfield  {title} {\bibinfo
  {title} {Probing the {{N{\'e}el Order}} in {{Altermagnetic
  RuO}}{\textsubscript{2}} {{Films}} via {{X-ray Magnetic Linear Dichroism}}},\
  }\href {https://doi.org/10.1088/0256-307X/42/2/027301} {\bibfield  {journal}
  {\bibinfo  {journal} {Chin. Phys. Lett.}\ }\textbf {\bibinfo {volume} {42}},\
  \bibinfo {pages} {027301} (\bibinfo {year} {2025}{\natexlab{b}})}\BibitemShut
  {NoStop}%
\bibitem [{\citenamefont {{Raboni Ferreira, Marina and Daurer, Benedikt J and
  Neethirajan, Jeffrey and Apseros, Andreas and Ruiz-G{\'o}mez, Sandra and
  Kaulich, Burkhard and Kazemian, Majid and Donnelly,
  Claire}}(2025)}]{raboni2025nanoscale}%
  \BibitemOpen
  \bibfield  {author} {\bibinfo {author} {\bibnamefont {{Raboni Ferreira,
  Marina and Daurer, Benedikt J and Neethirajan, Jeffrey and Apseros, Andreas
  and Ruiz-G{\'o}mez, Sandra and Kaulich, Burkhard and Kazemian, Majid and
  Donnelly, Claire}}},\ }\bibfield  {title} {\bibinfo {title} {{Nanoscale
  Mapping of Magnetic Orientations with Complex X-ray Magnetic Linear
  Dichroism}},\ }\href {https://arxiv.org/abs/2502.08617} {\bibfield  {journal}
  {\bibinfo  {journal} {arXiv:2502.08617}\ } (\bibinfo {year}
  {2025})}\BibitemShut {NoStop}%
\bibitem [{\citenamefont {Leivisk{\"a}}\ \emph {et~al.}(2024)\citenamefont
  {Leivisk{\"a}}, \citenamefont {Rial}, \citenamefont {Bad'ura}, \citenamefont
  {Seeger}, \citenamefont {Kounta}, \citenamefont {Beckert}, \citenamefont
  {Kriegner}, \citenamefont {Joumard}, \citenamefont {Schmoranzerov{\'a}},
  \citenamefont {Sinova}, \citenamefont {Gomonay}, \citenamefont {Thomas},
  \citenamefont {Goennenwein}, \citenamefont {Reichlov{\'a}}, \citenamefont
  {{\v S}mejkal}, \citenamefont {Michez}, \citenamefont {Jungwirth},\ and\
  \citenamefont {Baltz}}]{leiviska2024anisotropy}%
  \BibitemOpen
  \bibfield  {author} {\bibinfo {author} {\bibfnamefont {M.}~\bibnamefont
  {Leivisk{\"a}}}, \bibinfo {author} {\bibfnamefont {J.}~\bibnamefont {Rial}},
  \bibinfo {author} {\bibfnamefont {A.}~\bibnamefont {Bad'ura}}, \bibinfo
  {author} {\bibfnamefont {R.~L.}\ \bibnamefont {Seeger}}, \bibinfo {author}
  {\bibfnamefont {I.}~\bibnamefont {Kounta}}, \bibinfo {author} {\bibfnamefont
  {S.}~\bibnamefont {Beckert}}, \bibinfo {author} {\bibfnamefont
  {D.}~\bibnamefont {Kriegner}}, \bibinfo {author} {\bibfnamefont
  {I.}~\bibnamefont {Joumard}}, \bibinfo {author} {\bibfnamefont
  {E.}~\bibnamefont {Schmoranzerov{\'a}}}, \bibinfo {author} {\bibfnamefont
  {J.}~\bibnamefont {Sinova}}, \bibinfo {author} {\bibfnamefont
  {O.}~\bibnamefont {Gomonay}}, \bibinfo {author} {\bibfnamefont
  {A.}~\bibnamefont {Thomas}}, \bibinfo {author} {\bibfnamefont {S.~T.~B.}\
  \bibnamefont {Goennenwein}}, \bibinfo {author} {\bibfnamefont
  {H.}~\bibnamefont {Reichlov{\'a}}}, \bibinfo {author} {\bibfnamefont
  {L.}~\bibnamefont {{\v S}mejkal}}, \bibinfo {author} {\bibfnamefont
  {L.}~\bibnamefont {Michez}}, \bibinfo {author} {\bibfnamefont
  {T.}~\bibnamefont {Jungwirth}},\ and\ \bibinfo {author} {\bibfnamefont
  {V.}~\bibnamefont {Baltz}},\ }\bibfield  {title} {\bibinfo {title}
  {{Anisotropy of the Anomalous {{Hall}} Effect in Thin Films of the
  Altermagnet Candidate Mn$_5$Si$_3$}},\ }\href
  {https://doi.org/10.1103/PhysRevB.109.224430} {\bibfield  {journal} {\bibinfo
   {journal} {Phys. Rev. B}\ }\textbf {\bibinfo {volume} {109}},\ \bibinfo
  {pages} {224430} (\bibinfo {year} {2024})}\BibitemShut {NoStop}%
\bibitem [{\citenamefont {Zhou}\ \emph {et~al.}(2025)\citenamefont {Zhou},
  \citenamefont {Cheng}, \citenamefont {Hu}, \citenamefont {Chu}, \citenamefont
  {Bai}, \citenamefont {Han}, \citenamefont {Liu}, \citenamefont {Pan},\ and\
  \citenamefont {Song}}]{zhou2025manipulation}%
  \BibitemOpen
  \bibfield  {author} {\bibinfo {author} {\bibfnamefont {Z.}~\bibnamefont
  {Zhou}}, \bibinfo {author} {\bibfnamefont {X.}~\bibnamefont {Cheng}},
  \bibinfo {author} {\bibfnamefont {M.}~\bibnamefont {Hu}}, \bibinfo {author}
  {\bibfnamefont {R.}~\bibnamefont {Chu}}, \bibinfo {author} {\bibfnamefont
  {H.}~\bibnamefont {Bai}}, \bibinfo {author} {\bibfnamefont {L.}~\bibnamefont
  {Han}}, \bibinfo {author} {\bibfnamefont {J.}~\bibnamefont {Liu}}, \bibinfo
  {author} {\bibfnamefont {F.}~\bibnamefont {Pan}},\ and\ \bibinfo {author}
  {\bibfnamefont {C.}~\bibnamefont {Song}},\ }\bibfield  {title} {\bibinfo
  {title} {{Manipulation of the altermagnetic order in CrSb via crystal
  symmetry}},\ }\href {https://doi.org/10.1038/s41586-024-08436-3} {\bibfield
  {journal} {\bibinfo  {journal} {Nature}\ }\textbf {\bibinfo {volume} {638}},\
  \bibinfo {pages} {645} (\bibinfo {year} {2025})}\BibitemShut {NoStop}%
\bibitem [{\citenamefont {Wu}\ \emph {et~al.}(2025)\citenamefont {Wu},
  \citenamefont {Cheng}, \citenamefont {Wang}, \citenamefont {Zeng},
  \citenamefont {Liu},\ and\ \citenamefont {Li}}]{wu2025optical}%
  \BibitemOpen
  \bibfield  {author} {\bibinfo {author} {\bibfnamefont {A.}~\bibnamefont
  {Wu}}, \bibinfo {author} {\bibfnamefont {D.}~\bibnamefont {Cheng}}, \bibinfo
  {author} {\bibfnamefont {X.}~\bibnamefont {Wang}}, \bibinfo {author}
  {\bibfnamefont {M.}~\bibnamefont {Zeng}}, \bibinfo {author} {\bibfnamefont
  {C.}~\bibnamefont {Liu}},\ and\ \bibinfo {author} {\bibfnamefont
  {X.}~\bibnamefont {Li}},\ }\bibfield  {title} {\bibinfo {title} {{Optical
  signatures of noncentrosymmetric structural distortion in altermagnetic
  MnTe}},\ }\href {https://doi.org/10.48550/arXiv.2503.17742} {\bibfield
  {journal} {\bibinfo  {journal} {arXiv:2503.17742}\ } (\bibinfo {year}
  {2025})}\BibitemShut {NoStop}%
\bibitem [{\citenamefont {Han}\ \emph {et~al.}(2025)\citenamefont {Han},
  \citenamefont {Fu}, \citenamefont {Song}, \citenamefont {Zhu}, \citenamefont
  {Li}, \citenamefont {Zhu}, \citenamefont {Bai}, \citenamefont {Chu},
  \citenamefont {Dai}, \citenamefont {Liang} \emph
  {et~al.}}]{han2025discovery}%
  \BibitemOpen
  \bibfield  {author} {\bibinfo {author} {\bibfnamefont {L.}~\bibnamefont
  {Han}}, \bibinfo {author} {\bibfnamefont {X.}~\bibnamefont {Fu}}, \bibinfo
  {author} {\bibfnamefont {C.}~\bibnamefont {Song}}, \bibinfo {author}
  {\bibfnamefont {Y.}~\bibnamefont {Zhu}}, \bibinfo {author} {\bibfnamefont
  {X.}~\bibnamefont {Li}}, \bibinfo {author} {\bibfnamefont {Z.}~\bibnamefont
  {Zhu}}, \bibinfo {author} {\bibfnamefont {H.}~\bibnamefont {Bai}}, \bibinfo
  {author} {\bibfnamefont {R.}~\bibnamefont {Chu}}, \bibinfo {author}
  {\bibfnamefont {J.}~\bibnamefont {Dai}}, \bibinfo {author} {\bibfnamefont
  {S.}~\bibnamefont {Liang}}, \emph {et~al.},\ }\bibfield  {title} {\bibinfo
  {title} {Discovery of a large magnetic nonlinear hall effect in an
  altermagnet},\ }\href {https://doi.org/10.48550/arXiv.2502.04920} {\bibfield
  {journal} {\bibinfo  {journal} {arXiv:2502.04920}\ } (\bibinfo {year}
  {2025})}\BibitemShut {NoStop}%
\bibitem [{\citenamefont {Guo}\ \emph {et~al.}(2023)\citenamefont {Guo},
  \citenamefont {Gu}, \citenamefont {Gao},\ and\ \citenamefont
  {Lu}}]{guo2023altermagnetic}%
  \BibitemOpen
  \bibfield  {author} {\bibinfo {author} {\bibfnamefont {P.-J.}\ \bibnamefont
  {Guo}}, \bibinfo {author} {\bibfnamefont {Y.}~\bibnamefont {Gu}}, \bibinfo
  {author} {\bibfnamefont {Z.-F.}\ \bibnamefont {Gao}},\ and\ \bibinfo {author}
  {\bibfnamefont {Z.-Y.}\ \bibnamefont {Lu}},\ }\bibfield  {title} {\bibinfo
  {title} {{Altermagnetic ferroelectric LiFe$_2$F$_6$ and spin-triplet
  excitonic insulator phase}},\ }\href
  {https://doi.org/10.48550/arXiv.2312.13911} {\bibfield  {journal} {\bibinfo
  {journal} {arXiv:2312.13911}\ } (\bibinfo {year} {2023})}\BibitemShut
  {NoStop}%
\bibitem [{\citenamefont {Zhang}\ \emph {et~al.}(2024)\citenamefont {Zhang},
  \citenamefont {Cui}, \citenamefont {Li}, \citenamefont {Duan}, \citenamefont
  {Li}, \citenamefont {Yu},\ and\ \citenamefont {Yao}}]{zhang2024predictable}%
  \BibitemOpen
  \bibfield  {author} {\bibinfo {author} {\bibfnamefont {R.-W.}\ \bibnamefont
  {Zhang}}, \bibinfo {author} {\bibfnamefont {C.}~\bibnamefont {Cui}}, \bibinfo
  {author} {\bibfnamefont {R.}~\bibnamefont {Li}}, \bibinfo {author}
  {\bibfnamefont {J.}~\bibnamefont {Duan}}, \bibinfo {author} {\bibfnamefont
  {L.}~\bibnamefont {Li}}, \bibinfo {author} {\bibfnamefont {Z.-M.}\
  \bibnamefont {Yu}},\ and\ \bibinfo {author} {\bibfnamefont {Y.}~\bibnamefont
  {Yao}},\ }\bibfield  {title} {\bibinfo {title} {Predictable gate-field
  control of spin in altermagnets with spin-layer coupling},\ }\href
  {https://doi.org/10.1103/PhysRevLett.133.056401} {\bibfield  {journal}
  {\bibinfo  {journal} {Phys. Rev. Lett.}\ }\textbf {\bibinfo {volume} {133}},\
  \bibinfo {pages} {056401} (\bibinfo {year} {2024})}\BibitemShut {NoStop}%
\bibitem [{\citenamefont {Matsuda}\ \emph {et~al.}(2025)\citenamefont
  {Matsuda}, \citenamefont {Watanabe},\ and\ \citenamefont
  {Arita}}]{matsuda2024multiferroic}%
  \BibitemOpen
  \bibfield  {author} {\bibinfo {author} {\bibfnamefont {J.}~\bibnamefont
  {Matsuda}}, \bibinfo {author} {\bibfnamefont {H.}~\bibnamefont {Watanabe}},\
  and\ \bibinfo {author} {\bibfnamefont {R.}~\bibnamefont {Arita}},\ }\bibfield
   {title} {\bibinfo {title} {Multiferroic collinear antiferromagnets with
  hidden altermagnetic spin splitting},\ }\href
  {https://doi.org/10.1103/vgcs-bn8g} {\bibfield  {journal} {\bibinfo
  {journal} {Phys. Rev. Lett.}\ }\textbf {\bibinfo {volume} {134}},\ \bibinfo
  {pages} {226703} (\bibinfo {year} {2025})}\BibitemShut {NoStop}%
\bibitem [{\citenamefont {Sheng}\ \emph {et~al.}(2025)\citenamefont {Sheng},
  \citenamefont {Zhang}, \citenamefont {Liu},\ and\ \citenamefont
  {Wu}}]{sheng2024ubiquitous}%
  \BibitemOpen
  \bibfield  {author} {\bibinfo {author} {\bibfnamefont {Y.}~\bibnamefont
  {Sheng}}, \bibinfo {author} {\bibfnamefont {J.}~\bibnamefont {Zhang}},
  \bibinfo {author} {\bibfnamefont {J.}~\bibnamefont {Liu}},\ and\ \bibinfo
  {author} {\bibfnamefont {M.}~\bibnamefont {Wu}},\ }\bibfield  {title}
  {\bibinfo {title} {Ubiquitous van der waals altermagnetism with sliding/moire
  ferroelectricity},\ }\href {https://doi.org/10.1007/s11433-025-2698-5}
  {\bibfield  {journal} {\bibinfo  {journal} {Science China Physics, Mechanics
  \& Astronomy}\ }\textbf {\bibinfo {volume} {68}},\ \bibinfo {pages} {297511}
  (\bibinfo {year} {2025})}\BibitemShut {NoStop}%
\bibitem [{\citenamefont {Gu}\ \emph {et~al.}(2025)\citenamefont {Gu},
  \citenamefont {Liu}, \citenamefont {Zhu}, \citenamefont {Yananose},
  \citenamefont {Chen}, \citenamefont {Hu}, \citenamefont {Stroppa},\ and\
  \citenamefont {Liu}}]{gu2025ferroelectric}%
  \BibitemOpen
  \bibfield  {author} {\bibinfo {author} {\bibfnamefont {M.}~\bibnamefont
  {Gu}}, \bibinfo {author} {\bibfnamefont {Y.}~\bibnamefont {Liu}}, \bibinfo
  {author} {\bibfnamefont {H.}~\bibnamefont {Zhu}}, \bibinfo {author}
  {\bibfnamefont {K.}~\bibnamefont {Yananose}}, \bibinfo {author}
  {\bibfnamefont {X.}~\bibnamefont {Chen}}, \bibinfo {author} {\bibfnamefont
  {Y.}~\bibnamefont {Hu}}, \bibinfo {author} {\bibfnamefont {A.}~\bibnamefont
  {Stroppa}},\ and\ \bibinfo {author} {\bibfnamefont {Q.}~\bibnamefont {Liu}},\
  }\bibfield  {title} {\bibinfo {title} {Ferroelectric {{Switchable
  Altermagnetism}}},\ }\href {https://doi.org/10.1103/PhysRevLett.134.106802}
  {\bibfield  {journal} {\bibinfo  {journal} {Phys. Rev. Lett.}\ }\textbf
  {\bibinfo {volume} {134}},\ \bibinfo {pages} {106802} (\bibinfo {year}
  {2025})}\BibitemShut {NoStop}%
\bibitem [{\citenamefont {Duan}\ \emph {et~al.}(2025)\citenamefont {Duan},
  \citenamefont {Zhang}, \citenamefont {Zhu}, \citenamefont {Liu},
  \citenamefont {Zhang}, \citenamefont {{\v Z}uti{\'c}},\ and\ \citenamefont
  {Zhou}}]{duan2025antiferroelectric}%
  \BibitemOpen
  \bibfield  {author} {\bibinfo {author} {\bibfnamefont {X.}~\bibnamefont
  {Duan}}, \bibinfo {author} {\bibfnamefont {J.}~\bibnamefont {Zhang}},
  \bibinfo {author} {\bibfnamefont {Z.}~\bibnamefont {Zhu}}, \bibinfo {author}
  {\bibfnamefont {Y.}~\bibnamefont {Liu}}, \bibinfo {author} {\bibfnamefont
  {Z.}~\bibnamefont {Zhang}}, \bibinfo {author} {\bibfnamefont
  {I.}~\bibnamefont {{\v Z}uti{\'c}}},\ and\ \bibinfo {author} {\bibfnamefont
  {T.}~\bibnamefont {Zhou}},\ }\bibfield  {title} {\bibinfo {title}
  {Antiferroelectric {{Altermagnets}}: {{Antiferroelectricity Alters
  Magnets}}},\ }\href {https://doi.org/10.1103/PhysRevLett.134.106801}
  {\bibfield  {journal} {\bibinfo  {journal} {Phys. Rev. Lett.}\ }\textbf
  {\bibinfo {volume} {134}},\ \bibinfo {pages} {106801} (\bibinfo {year}
  {2025})}\BibitemShut {NoStop}%
\bibitem [{\citenamefont {Camerano}\ \emph {et~al.}(2025)\citenamefont
  {Camerano}, \citenamefont {Fumega}, \citenamefont {Lado}, \citenamefont
  {Stroppa},\ and\ \citenamefont {Profeta}}]{camerano2025multiferroic}%
  \BibitemOpen
  \bibfield  {author} {\bibinfo {author} {\bibfnamefont {L.}~\bibnamefont
  {Camerano}}, \bibinfo {author} {\bibfnamefont {A.~O.}\ \bibnamefont
  {Fumega}}, \bibinfo {author} {\bibfnamefont {J.~L.}\ \bibnamefont {Lado}},
  \bibinfo {author} {\bibfnamefont {A.}~\bibnamefont {Stroppa}},\ and\ \bibinfo
  {author} {\bibfnamefont {G.}~\bibnamefont {Profeta}},\ }\bibfield  {title}
  {\bibinfo {title} {Multiferroic nematic d-wave altermagnetism driven by
  orbital-order on the honeycomb lattice},\ }\href
  {https://doi.org/10.1038/s41699-025-00599-5} {\bibfield  {journal} {\bibinfo
  {journal} {npj 2D Materials and Applications}\ }\textbf {\bibinfo {volume}
  {9}},\ \bibinfo {pages} {75} (\bibinfo {year} {2025})}\BibitemShut {NoStop}%
\bibitem [{\citenamefont {Wang}\ \emph
  {et~al.}(2025{\natexlab{b}})\citenamefont {Wang}, \citenamefont {Wang},
  \citenamefont {Fan}, \citenamefont {Zhou}, \citenamefont {Li},\ and\
  \citenamefont {Wang}}]{wang2025two}%
  \BibitemOpen
  \bibfield  {author} {\bibinfo {author} {\bibfnamefont {S.}~\bibnamefont
  {Wang}}, \bibinfo {author} {\bibfnamefont {W.-W.}\ \bibnamefont {Wang}},
  \bibinfo {author} {\bibfnamefont {J.}~\bibnamefont {Fan}}, \bibinfo {author}
  {\bibfnamefont {X.}~\bibnamefont {Zhou}}, \bibinfo {author} {\bibfnamefont
  {X.-P.}\ \bibnamefont {Li}},\ and\ \bibinfo {author} {\bibfnamefont
  {L.}~\bibnamefont {Wang}},\ }\bibfield  {title} {\bibinfo {title}
  {Two-dimensional dual-switchable ferroelectric altermagnets: Altering
  electrons and magnons},\ }\href
  {https://doi.org/10.1021/acs.nanolett.5c03483} {\bibfield  {journal}
  {\bibinfo  {journal} {Nano Letters}\ }\textbf {\bibinfo {volume} {25}},\
  \bibinfo {pages} {14618} (\bibinfo {year} {2025}{\natexlab{b}})}\BibitemShut
  {NoStop}%
\bibitem [{\citenamefont {Dong}\ \emph {et~al.}(2019)\citenamefont {Dong},
  \citenamefont {Xiang},\ and\ \citenamefont
  {Dagotto}}]{dong2019magnetoelectricity}%
  \BibitemOpen
  \bibfield  {author} {\bibinfo {author} {\bibfnamefont {S.}~\bibnamefont
  {Dong}}, \bibinfo {author} {\bibfnamefont {H.}~\bibnamefont {Xiang}},\ and\
  \bibinfo {author} {\bibfnamefont {E.}~\bibnamefont {Dagotto}},\ }\bibfield
  {title} {\bibinfo {title} {Magnetoelectricity in multiferroics: A theoretical
  perspective},\ }\href {https://doi.org/10.1093/nsr/nwz023} {\bibfield
  {journal} {\bibinfo  {journal} {Nat. Sci. Rev.}\ }\textbf {\bibinfo {volume}
  {6}},\ \bibinfo {pages} {629} (\bibinfo {year} {2019})}\BibitemShut {NoStop}%
\bibitem [{\citenamefont {Li}\ \emph {et~al.}(2023)\citenamefont {Li},
  \citenamefont {Xu}, \citenamefont {Liu}, \citenamefont {Li}, \citenamefont
  {Bellaiche},\ and\ \citenamefont {Xiang}}]{li2023realistic}%
  \BibitemOpen
  \bibfield  {author} {\bibinfo {author} {\bibfnamefont {X.}~\bibnamefont
  {Li}}, \bibinfo {author} {\bibfnamefont {C.}~\bibnamefont {Xu}}, \bibinfo
  {author} {\bibfnamefont {B.}~\bibnamefont {Liu}}, \bibinfo {author}
  {\bibfnamefont {X.}~\bibnamefont {Li}}, \bibinfo {author} {\bibfnamefont
  {L.}~\bibnamefont {Bellaiche}},\ and\ \bibinfo {author} {\bibfnamefont
  {H.}~\bibnamefont {Xiang}},\ }\bibfield  {title} {\bibinfo {title}
  {{Realistic Spin Model for Multiferroic ${\mathrm{NiI}}_{2}$}},\ }\href
  {https://doi.org/10.1103/PhysRevLett.131.036701} {\bibfield  {journal}
  {\bibinfo  {journal} {Phys. Rev. Lett.}\ }\textbf {\bibinfo {volume} {131}},\
  \bibinfo {pages} {036701} (\bibinfo {year} {2023})}\BibitemShut {NoStop}%
\bibitem [{\citenamefont {Liu}\ \emph {et~al.}(2024{\natexlab{b}})\citenamefont
  {Liu}, \citenamefont {Ren},\ and\ \citenamefont {Picozzi}}]{liu2024spin}%
  \BibitemOpen
  \bibfield  {author} {\bibinfo {author} {\bibfnamefont {C.}~\bibnamefont
  {Liu}}, \bibinfo {author} {\bibfnamefont {W.}~\bibnamefont {Ren}},\ and\
  \bibinfo {author} {\bibfnamefont {S.}~\bibnamefont {Picozzi}},\ }\bibfield
  {title} {\bibinfo {title} {Spin-chirality-driven multiferroicity in van der
  waals monolayers},\ }\href {https://doi.org/10.1103/PhysRevLett.132.086802}
  {\bibfield  {journal} {\bibinfo  {journal} {Phys. Rev. Lett.}\ }\textbf
  {\bibinfo {volume} {132}},\ \bibinfo {pages} {086802} (\bibinfo {year}
  {2024}{\natexlab{b}})}\BibitemShut {NoStop}%
\bibitem [{\citenamefont {Stavri\ifmmode~\acute{c}\else \'{c}\fi{}}\ \emph
  {et~al.}(2025)\citenamefont {Stavri\ifmmode~\acute{c}\else \'{c}\fi{}},
  \citenamefont {Cuono}, \citenamefont {Yang}, \citenamefont {Puente-Uriona},
  \citenamefont {Iba\~nez Azpiroz}, \citenamefont {Barone}, \citenamefont
  {Droghetti},\ and\ \citenamefont {Picozzi}}]{stavric2025giant}%
  \BibitemOpen
  \bibfield  {author} {\bibinfo {author} {\bibfnamefont {S.}~\bibnamefont
  {Stavri\ifmmode~\acute{c}\else \'{c}\fi{}}}, \bibinfo {author} {\bibfnamefont
  {G.}~\bibnamefont {Cuono}}, \bibinfo {author} {\bibfnamefont
  {B.}~\bibnamefont {Yang}}, \bibinfo {author} {\bibfnamefont {A.~R.}\
  \bibnamefont {Puente-Uriona}}, \bibinfo {author} {\bibfnamefont
  {J.}~\bibnamefont {Iba\~nez Azpiroz}}, \bibinfo {author} {\bibfnamefont
  {P.}~\bibnamefont {Barone}}, \bibinfo {author} {\bibfnamefont
  {A.}~\bibnamefont {Droghetti}},\ and\ \bibinfo {author} {\bibfnamefont
  {S.}~\bibnamefont {Picozzi}},\ }\bibfield  {title} {\bibinfo {title} {Giant
  nonreciprocal band structure effect in a multiferroic material},\ }\href
  {https://doi.org/10.1103/15ws-ftbf} {\bibfield  {journal} {\bibinfo
  {journal} {Phys. Rev. Lett.}\ }\textbf {\bibinfo {volume} {135}},\ \bibinfo
  {pages} {206401} (\bibinfo {year} {2025})}\BibitemShut {NoStop}%
\bibitem [{\citenamefont {Murakawa}\ \emph {et~al.}(2010)\citenamefont
  {Murakawa}, \citenamefont {Onose}, \citenamefont {Miyahara}, \citenamefont
  {Furukawa},\ and\ \citenamefont {Tokura}}]{murakawa2010ferroelectricity}%
  \BibitemOpen
  \bibfield  {author} {\bibinfo {author} {\bibfnamefont {H.}~\bibnamefont
  {Murakawa}}, \bibinfo {author} {\bibfnamefont {Y.}~\bibnamefont {Onose}},
  \bibinfo {author} {\bibfnamefont {S.}~\bibnamefont {Miyahara}}, \bibinfo
  {author} {\bibfnamefont {N.}~\bibnamefont {Furukawa}},\ and\ \bibinfo
  {author} {\bibfnamefont {Y.}~\bibnamefont {Tokura}},\ }\bibfield  {title}
  {\bibinfo {title} {{Ferroelectricity induced by spin-dependent metal-ligand
  hybridization in Ba$_2$CoGe$_2$O$_7$}},\ }\href
  {https://doi.org/10.1103/PhysRevLett.105.137202} {\bibfield  {journal}
  {\bibinfo  {journal} {Phys. Rev. Lett.}\ }\textbf {\bibinfo {volume} {105}},\
  \bibinfo {pages} {137202} (\bibinfo {year} {2010})}\BibitemShut {NoStop}%
\bibitem [{\citenamefont {Zhang}\ \emph {et~al.}(2022)\citenamefont {Zhang},
  \citenamefont {Zhou}, \citenamefont {Wang}, \citenamefont {Shen},
  \citenamefont {Wang},\ and\ \citenamefont {Lu}}]{zhang2022coexistence}%
  \BibitemOpen
  \bibfield  {author} {\bibinfo {author} {\bibfnamefont {J.}~\bibnamefont
  {Zhang}}, \bibinfo {author} {\bibfnamefont {Y.}~\bibnamefont {Zhou}},
  \bibinfo {author} {\bibfnamefont {F.}~\bibnamefont {Wang}}, \bibinfo {author}
  {\bibfnamefont {X.}~\bibnamefont {Shen}}, \bibinfo {author} {\bibfnamefont
  {J.}~\bibnamefont {Wang}},\ and\ \bibinfo {author} {\bibfnamefont
  {X.}~\bibnamefont {Lu}},\ }\bibfield  {title} {\bibinfo {title} {Coexistence
  and coupling of spin-induced ferroelectricity and ferromagnetism in
  perovskites},\ }\href {https://doi.org/10.1103/PhysRevLett.129.117603}
  {\bibfield  {journal} {\bibinfo  {journal} {Phys. Rev. Lett.}\ }\textbf
  {\bibinfo {volume} {129}},\ \bibinfo {pages} {117603} (\bibinfo {year}
  {2022})}\BibitemShut {NoStop}%
\bibitem [{\citenamefont {Zhou}\ \emph
  {et~al.}(2024{\natexlab{b}})\citenamefont {Zhou}, \citenamefont {Ye},
  \citenamefont {Zhang},\ and\ \citenamefont {Dong}}]{zhou2024double}%
  \BibitemOpen
  \bibfield  {author} {\bibinfo {author} {\bibfnamefont {Y.}~\bibnamefont
  {Zhou}}, \bibinfo {author} {\bibfnamefont {H.}~\bibnamefont {Ye}}, \bibinfo
  {author} {\bibfnamefont {J.}~\bibnamefont {Zhang}},\ and\ \bibinfo {author}
  {\bibfnamefont {S.}~\bibnamefont {Dong}},\ }\bibfield  {title} {\bibinfo
  {title} {Double-leaf {{Riemann}} surface topological converse
  magnetoelectricity},\ }\href {https://doi.org/10.1103/PhysRevB.110.054424}
  {\bibfield  {journal} {\bibinfo  {journal} {Phys. Rev. B}\ }\textbf {\bibinfo
  {volume} {110}},\ \bibinfo {pages} {054424} (\bibinfo {year}
  {2024}{\natexlab{b}})}\BibitemShut {NoStop}%
\bibitem [{\citenamefont {Jia}\ \emph {et~al.}(2006)\citenamefont {Jia},
  \citenamefont {Onoda}, \citenamefont {Nagaosa},\ and\ \citenamefont
  {Han}}]{jia2006bond}%
  \BibitemOpen
  \bibfield  {author} {\bibinfo {author} {\bibfnamefont {C.}~\bibnamefont
  {Jia}}, \bibinfo {author} {\bibfnamefont {S.}~\bibnamefont {Onoda}}, \bibinfo
  {author} {\bibfnamefont {N.}~\bibnamefont {Nagaosa}},\ and\ \bibinfo {author}
  {\bibfnamefont {J.~H.}\ \bibnamefont {Han}},\ }\bibfield  {title} {\bibinfo
  {title} {Bond electronic polarization induced by spin},\ }\href
  {https://doi.org/10.1103/PhysRevB.74.224444} {\bibfield  {journal} {\bibinfo
  {journal} {Phys. Rev. B}\ }\textbf {\bibinfo {volume} {74}},\ \bibinfo
  {pages} {224444} (\bibinfo {year} {2006})}\BibitemShut {NoStop}%
\bibitem [{\citenamefont {Matsumoto}\ \emph {et~al.}(2017)\citenamefont
  {Matsumoto}, \citenamefont {Chimata},\ and\ \citenamefont
  {Koga}}]{matsumoto2017symmetry}%
  \BibitemOpen
  \bibfield  {author} {\bibinfo {author} {\bibfnamefont {M.}~\bibnamefont
  {Matsumoto}}, \bibinfo {author} {\bibfnamefont {K.}~\bibnamefont {Chimata}},\
  and\ \bibinfo {author} {\bibfnamefont {M.}~\bibnamefont {Koga}},\ }\bibfield
  {title} {\bibinfo {title} {Symmetry analysis of spin-dependent electric
  dipole and its application to magnetoelectric effects},\ }\href
  {https://doi.org/10.7566/JPSJ.86.034704} {\bibfield  {journal} {\bibinfo
  {journal} {J. Phys. Soc. Jpn.}\ }\textbf {\bibinfo {volume} {86}},\ \bibinfo
  {pages} {034704} (\bibinfo {year} {2017})}\BibitemShut {NoStop}%
\bibitem [{\citenamefont {Wu}\ \emph {et~al.}(2024)\citenamefont {Wu},
  \citenamefont {Zeng}, \citenamefont {Lu}, \citenamefont {Han}, \citenamefont
  {Yang}, \citenamefont {Liu}, \citenamefont {Zhao}, \citenamefont {Qiao},
  \citenamefont {Ji}, \citenamefont {Che} \emph {et~al.}}]{wu2024coexistence}%
  \BibitemOpen
  \bibfield  {author} {\bibinfo {author} {\bibfnamefont {Y.}~\bibnamefont
  {Wu}}, \bibinfo {author} {\bibfnamefont {Z.}~\bibnamefont {Zeng}}, \bibinfo
  {author} {\bibfnamefont {H.}~\bibnamefont {Lu}}, \bibinfo {author}
  {\bibfnamefont {X.}~\bibnamefont {Han}}, \bibinfo {author} {\bibfnamefont
  {C.}~\bibnamefont {Yang}}, \bibinfo {author} {\bibfnamefont {N.}~\bibnamefont
  {Liu}}, \bibinfo {author} {\bibfnamefont {X.}~\bibnamefont {Zhao}}, \bibinfo
  {author} {\bibfnamefont {L.}~\bibnamefont {Qiao}}, \bibinfo {author}
  {\bibfnamefont {W.}~\bibnamefont {Ji}}, \bibinfo {author} {\bibfnamefont
  {R.}~\bibnamefont {Che}}, \emph {et~al.},\ }\bibfield  {title} {\bibinfo
  {title} {Coexistence of ferroelectricity and antiferroelectricity in 2d van
  der waals multiferroic},\ }\href {https://doi.org/10.1038/s41467-024-53019-5}
  {\bibfield  {journal} {\bibinfo  {journal} {Nat. Commun.}\ }\textbf {\bibinfo
  {volume} {15}},\ \bibinfo {pages} {8616} (\bibinfo {year}
  {2024})}\BibitemShut {NoStop}%
\bibitem [{\citenamefont {Matsukura}\ \emph {et~al.}(2015)\citenamefont
  {Matsukura}, \citenamefont {Tokura},\ and\ \citenamefont
  {Ohno}}]{matsukura2015control}%
  \BibitemOpen
  \bibfield  {author} {\bibinfo {author} {\bibfnamefont {F.}~\bibnamefont
  {Matsukura}}, \bibinfo {author} {\bibfnamefont {Y.}~\bibnamefont {Tokura}},\
  and\ \bibinfo {author} {\bibfnamefont {H.}~\bibnamefont {Ohno}},\ }\bibfield
  {title} {\bibinfo {title} {Control of magnetism by electric fields},\ }\href
  {https://doi.org/10.1038/nnano.2015.22} {\bibfield  {journal} {\bibinfo
  {journal} {Nat. Nanotechnol.}\ }\textbf {\bibinfo {volume} {10}},\ \bibinfo
  {pages} {209} (\bibinfo {year} {2015})}\BibitemShut {NoStop}%
\bibitem [{\citenamefont {Jiang}\ \emph {et~al.}(2018)\citenamefont {Jiang},
  \citenamefont {Li}, \citenamefont {Wang}, \citenamefont {Mak},\ and\
  \citenamefont {Shan}}]{jiang2018controlling}%
  \BibitemOpen
  \bibfield  {author} {\bibinfo {author} {\bibfnamefont {S.}~\bibnamefont
  {Jiang}}, \bibinfo {author} {\bibfnamefont {L.}~\bibnamefont {Li}}, \bibinfo
  {author} {\bibfnamefont {Z.}~\bibnamefont {Wang}}, \bibinfo {author}
  {\bibfnamefont {K.~F.}\ \bibnamefont {Mak}},\ and\ \bibinfo {author}
  {\bibfnamefont {J.}~\bibnamefont {Shan}},\ }\bibfield  {title} {\bibinfo
  {title} {{Controlling Magnetism in 2D CrI$_3$ by Electrostatic Doping}},\
  }\href {https://doi.org/10.1038/s41565-018-0135-x} {\bibfield  {journal}
  {\bibinfo  {journal} {Nat. Nanotechnol.}\ }\textbf {\bibinfo {volume} {13}},\
  \bibinfo {pages} {549} (\bibinfo {year} {2018})}\BibitemShut {NoStop}%
\bibitem [{\citenamefont {Song}\ \emph {et~al.}(2022)\citenamefont {Song},
  \citenamefont {Occhialini}, \citenamefont {Erge{\c c}en}, \citenamefont
  {Ilyas}, \citenamefont {Amoroso}, \citenamefont {Barone}, \citenamefont
  {Kapeghian}, \citenamefont {Watanabe}, \citenamefont {Taniguchi},
  \citenamefont {Botana}, \citenamefont {Picozzi}, \citenamefont {Gedik},\ and\
  \citenamefont {Comin}}]{song2022evidence}%
  \BibitemOpen
  \bibfield  {author} {\bibinfo {author} {\bibfnamefont {Q.}~\bibnamefont
  {Song}}, \bibinfo {author} {\bibfnamefont {C.~A.}\ \bibnamefont
  {Occhialini}}, \bibinfo {author} {\bibfnamefont {E.}~\bibnamefont {Erge{\c
  c}en}}, \bibinfo {author} {\bibfnamefont {B.}~\bibnamefont {Ilyas}}, \bibinfo
  {author} {\bibfnamefont {D.}~\bibnamefont {Amoroso}}, \bibinfo {author}
  {\bibfnamefont {P.}~\bibnamefont {Barone}}, \bibinfo {author} {\bibfnamefont
  {J.}~\bibnamefont {Kapeghian}}, \bibinfo {author} {\bibfnamefont
  {K.}~\bibnamefont {Watanabe}}, \bibinfo {author} {\bibfnamefont
  {T.}~\bibnamefont {Taniguchi}}, \bibinfo {author} {\bibfnamefont {A.~S.}\
  \bibnamefont {Botana}}, \bibinfo {author} {\bibfnamefont {S.}~\bibnamefont
  {Picozzi}}, \bibinfo {author} {\bibfnamefont {N.}~\bibnamefont {Gedik}},\
  and\ \bibinfo {author} {\bibfnamefont {R.}~\bibnamefont {Comin}},\ }\bibfield
   {title} {\bibinfo {title} {Evidence for a single-layer van der {{Waals}}
  multiferroic},\ }\href {https://doi.org/10.1038/s41586-021-04337-x}
  {\bibfield  {journal} {\bibinfo  {journal} {Nature}\ }\textbf {\bibinfo
  {volume} {602}},\ \bibinfo {pages} {601} (\bibinfo {year}
  {2022})}\BibitemShut {NoStop}%
\bibitem [{\citenamefont {Murakawa}\ \emph {et~al.}(2012)\citenamefont
  {Murakawa}, \citenamefont {Onose}, \citenamefont {Miyahara}, \citenamefont
  {Furukawa},\ and\ \citenamefont {Tokura}}]{murakawa2012comprehensive}%
  \BibitemOpen
  \bibfield  {author} {\bibinfo {author} {\bibfnamefont {H.}~\bibnamefont
  {Murakawa}}, \bibinfo {author} {\bibfnamefont {Y.}~\bibnamefont {Onose}},
  \bibinfo {author} {\bibfnamefont {S.}~\bibnamefont {Miyahara}}, \bibinfo
  {author} {\bibfnamefont {N.}~\bibnamefont {Furukawa}},\ and\ \bibinfo
  {author} {\bibfnamefont {Y.}~\bibnamefont {Tokura}},\ }\bibfield  {title}
  {\bibinfo {title} {{Comprehensive study of the ferroelectricity induced by
  the spin-dependent $d$-$p$ hybridization mechanism in
  Ba${}_{2}X$Ge${}_{2}$O${}_{7}$ ($X$ = Mn, Co, and Cu)}},\ }\href
  {https://doi.org/10.1103/PhysRevB.85.174106} {\bibfield  {journal} {\bibinfo
  {journal} {Phys. Rev. B}\ }\textbf {\bibinfo {volume} {85}},\ \bibinfo
  {pages} {174106} (\bibinfo {year} {2012})}\BibitemShut {NoStop}%
\bibitem [{\citenamefont {Jia}\ \emph {et~al.}(2007)\citenamefont {Jia},
  \citenamefont {Onoda}, \citenamefont {Nagaosa},\ and\ \citenamefont
  {Han}}]{jia2007Microscopic}%
  \BibitemOpen
  \bibfield  {author} {\bibinfo {author} {\bibfnamefont {C.}~\bibnamefont
  {Jia}}, \bibinfo {author} {\bibfnamefont {S.}~\bibnamefont {Onoda}}, \bibinfo
  {author} {\bibfnamefont {N.}~\bibnamefont {Nagaosa}},\ and\ \bibinfo {author}
  {\bibfnamefont {J.~H.}\ \bibnamefont {Han}},\ }\bibfield  {title} {\bibinfo
  {title} {Microscopic theory of spin-polarization coupling in multiferroic
  transition metal oxides},\ }\href
  {https://doi.org/10.1103/PhysRevB.76.144424} {\bibfield  {journal} {\bibinfo
  {journal} {Phys. Rev. B}\ }\textbf {\bibinfo {volume} {76}},\ \bibinfo
  {pages} {144424} (\bibinfo {year} {2007})}\BibitemShut {NoStop}%
\bibitem [{\citenamefont {Behovits}\ \emph {et~al.}(2023)\citenamefont
  {Behovits}, \citenamefont {Chekhov}, \citenamefont {Bodnar}, \citenamefont
  {Gueckstock}, \citenamefont {Reimers}, \citenamefont {Lytvynenko},
  \citenamefont {Skourski}, \citenamefont {Wolf}, \citenamefont {Seifert},
  \citenamefont {Gomonay}, \citenamefont {Kläui}, \citenamefont {Jourdan},\
  and\ \citenamefont {Kampfrath}}]{behovits2023Terahertz}%
  \BibitemOpen
  \bibfield  {author} {\bibinfo {author} {\bibfnamefont {Y.}~\bibnamefont
  {Behovits}}, \bibinfo {author} {\bibfnamefont {A.~L.}\ \bibnamefont
  {Chekhov}}, \bibinfo {author} {\bibfnamefont {S.~Y.}\ \bibnamefont {Bodnar}},
  \bibinfo {author} {\bibfnamefont {O.}~\bibnamefont {Gueckstock}}, \bibinfo
  {author} {\bibfnamefont {S.}~\bibnamefont {Reimers}}, \bibinfo {author}
  {\bibfnamefont {Y.}~\bibnamefont {Lytvynenko}}, \bibinfo {author}
  {\bibfnamefont {Y.}~\bibnamefont {Skourski}}, \bibinfo {author}
  {\bibfnamefont {M.}~\bibnamefont {Wolf}}, \bibinfo {author} {\bibfnamefont
  {T.~S.}\ \bibnamefont {Seifert}}, \bibinfo {author} {\bibfnamefont
  {O.}~\bibnamefont {Gomonay}}, \bibinfo {author} {\bibfnamefont
  {M.}~\bibnamefont {Kläui}}, \bibinfo {author} {\bibfnamefont
  {M.}~\bibnamefont {Jourdan}},\ and\ \bibinfo {author} {\bibfnamefont
  {T.}~\bibnamefont {Kampfrath}},\ }\bibfield  {title} {\bibinfo {title}
  {Terahertz {{Néel}} spin-orbit torques drive nonlinear magnon dynamics in
  antiferromagnetic {{Mn2Au}}},\ }\href
  {https://doi.org/10.1038/s41467-023-41569-z} {\bibfield  {journal} {\bibinfo
  {journal} {Nat. Commun.}\ }\textbf {\bibinfo {volume} {14}},\ \bibinfo
  {pages} {6038} (\bibinfo {year} {2023})}\BibitemShut {NoStop}%
\bibitem [{\citenamefont {Sun}\ \emph {et~al.}(2025)\citenamefont {Sun},
  \citenamefont {Yang}, \citenamefont {Wang}, \citenamefont {Liu},
  \citenamefont {Wang}, \citenamefont {Huang},\ and\ \citenamefont
  {Cheng}}]{sun2025proposing}%
  \BibitemOpen
  \bibfield  {author} {\bibinfo {author} {\bibfnamefont {W.}~\bibnamefont
  {Sun}}, \bibinfo {author} {\bibfnamefont {C.}~\bibnamefont {Yang}}, \bibinfo
  {author} {\bibfnamefont {W.}~\bibnamefont {Wang}}, \bibinfo {author}
  {\bibfnamefont {Y.}~\bibnamefont {Liu}}, \bibinfo {author} {\bibfnamefont
  {X.}~\bibnamefont {Wang}}, \bibinfo {author} {\bibfnamefont {S.}~\bibnamefont
  {Huang}},\ and\ \bibinfo {author} {\bibfnamefont {Z.}~\bibnamefont {Cheng}},\
  }\bibfield  {title} {\bibinfo {title} {{Proposing Altermagnetic-Ferroelectric
  Type-III Multiferroics with Robust Magnetoelectric Coupling}},\ }\href
  {https://doi.org/10.1002/adma.202502575} {\bibfield  {journal} {\bibinfo
  {journal} {Adv. Mater.}\ ,\ \bibinfo {pages} {2502575}} (\bibinfo {year}
  {2025})}\BibitemShut {NoStop}%
\bibitem [{\citenamefont {{\v S}mejkal}(2024)}]{vsmejkal2024altermagnetic}%
  \BibitemOpen
  \bibfield  {author} {\bibinfo {author} {\bibfnamefont {L.}~\bibnamefont {{\v
  S}mejkal}},\ }\bibfield  {title} {\bibinfo {title} {Altermagnetic
  multiferroics and altermagnetoelectric effect},\ }\href
  {https://doi.org/10.48550/arXiv.2411.19928} {\bibfield  {journal} {\bibinfo
  {journal} {arXiv:2411.19928}\ } (\bibinfo {year} {2024})}\BibitemShut
  {NoStop}%
\bibitem [{\citenamefont {Cao}\ \emph {et~al.}(2024)\citenamefont {Cao},
  \citenamefont {Dong}, \citenamefont {Fei},\ and\ \citenamefont
  {Yao}}]{cao2024designing}%
  \BibitemOpen
  \bibfield  {author} {\bibinfo {author} {\bibfnamefont {R.}~\bibnamefont
  {Cao}}, \bibinfo {author} {\bibfnamefont {R.}~\bibnamefont {Dong}}, \bibinfo
  {author} {\bibfnamefont {R.}~\bibnamefont {Fei}},\ and\ \bibinfo {author}
  {\bibfnamefont {Y.}~\bibnamefont {Yao}},\ }\bibfield  {title} {\bibinfo
  {title} {Designing spin-driven multiferroics in altermagnets},\ }\href
  {https://arxiv.org/abs/2412.20347} {\bibfield  {journal} {\bibinfo  {journal}
  {arXiv:2412.20347}\ } (\bibinfo {year} {2024})}\BibitemShut {NoStop}%
\bibitem [{\citenamefont {Zhu}\ \emph {et~al.}(2025{\natexlab{a}})\citenamefont
  {Zhu}, \citenamefont {Duan}, \citenamefont {Zhang}, \citenamefont {Hao},
  \citenamefont {Zutic},\ and\ \citenamefont {Zhou}}]{zhu2025two}%
  \BibitemOpen
  \bibfield  {author} {\bibinfo {author} {\bibfnamefont {Z.}~\bibnamefont
  {Zhu}}, \bibinfo {author} {\bibfnamefont {X.}~\bibnamefont {Duan}}, \bibinfo
  {author} {\bibfnamefont {J.}~\bibnamefont {Zhang}}, \bibinfo {author}
  {\bibfnamefont {B.}~\bibnamefont {Hao}}, \bibinfo {author} {\bibfnamefont
  {I.}~\bibnamefont {Zutic}},\ and\ \bibinfo {author} {\bibfnamefont
  {T.}~\bibnamefont {Zhou}},\ }\bibfield  {title} {\bibinfo {title}
  {Two-dimensional ferroelectric altermagnets: From model to material
  realization},\ }\href {https://doi.org/10.1021/acs.nanolett.5c02121}
  {\bibfield  {journal} {\bibinfo  {journal} {Nano Lett.}\ } (\bibinfo {year}
  {2025}{\natexlab{a}})}\BibitemShut {NoStop}%
\bibitem [{\citenamefont {Zhu}\ \emph {et~al.}(2025{\natexlab{b}})\citenamefont
  {Zhu}, \citenamefont {Gu}, \citenamefont {Liu}, \citenamefont {Chen},
  \citenamefont {Li}, \citenamefont {Du},\ and\ \citenamefont
  {Liu}}]{zhu2025sliding}%
  \BibitemOpen
  \bibfield  {author} {\bibinfo {author} {\bibfnamefont {Y.}~\bibnamefont
  {Zhu}}, \bibinfo {author} {\bibfnamefont {M.}~\bibnamefont {Gu}}, \bibinfo
  {author} {\bibfnamefont {Y.}~\bibnamefont {Liu}}, \bibinfo {author}
  {\bibfnamefont {X.}~\bibnamefont {Chen}}, \bibinfo {author} {\bibfnamefont
  {Y.}~\bibnamefont {Li}}, \bibinfo {author} {\bibfnamefont {S.}~\bibnamefont
  {Du}},\ and\ \bibinfo {author} {\bibfnamefont {Q.}~\bibnamefont {Liu}},\
  }\bibfield  {title} {\bibinfo {title} {Sliding ferroelectric control of
  unconventional magnetism in stacked bilayers},\ }\href
  {https://doi.org/10.1103/dmzg-ck2t} {\bibfield  {journal} {\bibinfo
  {journal} {Phys. Rev. Lett.}\ }\textbf {\bibinfo {volume} {135}},\ \bibinfo
  {pages} {056801} (\bibinfo {year} {2025}{\natexlab{b}})}\BibitemShut
  {NoStop}%
\bibitem [{\citenamefont {Zhu}\ \emph {et~al.}(2025{\natexlab{c}})\citenamefont
  {Zhu}, \citenamefont {Liu}, \citenamefont {Duan}, \citenamefont {Zhang},
  \citenamefont {Hao}, \citenamefont {Wei}, \citenamefont {Žutić},\ and\
  \citenamefont {Zhou}}]{zhu2025emergent}%
  \BibitemOpen
  \bibfield  {author} {\bibinfo {author} {\bibfnamefont {Z.}~\bibnamefont
  {Zhu}}, \bibinfo {author} {\bibfnamefont {Y.}~\bibnamefont {Liu}}, \bibinfo
  {author} {\bibfnamefont {X.}~\bibnamefont {Duan}}, \bibinfo {author}
  {\bibfnamefont {J.}~\bibnamefont {Zhang}}, \bibinfo {author} {\bibfnamefont
  {B.}~\bibnamefont {Hao}}, \bibinfo {author} {\bibfnamefont {S.-H.}\
  \bibnamefont {Wei}}, \bibinfo {author} {\bibfnamefont {I.}~\bibnamefont
  {Žutić}},\ and\ \bibinfo {author} {\bibfnamefont {T.}~\bibnamefont
  {Zhou}},\ }\bibfield  {title} {\bibinfo {title} {Emergent multiferroic
  altermagnets and spin control via noncollinear molecular polarization},\
  }\href {https://doi.org/10.1007/s11433-025-2778-3} {\bibfield  {journal}
  {\bibinfo  {journal} {Sci. China Phys. Mech.}\ }\textbf {\bibinfo {volume}
  {68}},\ \bibinfo {pages} {127562} (\bibinfo {year}
  {2025}{\natexlab{c}})}\BibitemShut {NoStop}%
\bibitem [{\citenamefont {Guo}\ \emph {et~al.}(2025)\citenamefont {Guo},
  \citenamefont {Dai}, \citenamefont {Wang}, \citenamefont {Wang},\ and\
  \citenamefont {Ji}}]{guo2025Mechanically}%
  \BibitemOpen
  \bibfield  {author} {\bibinfo {author} {\bibfnamefont {D.}~\bibnamefont
  {Guo}}, \bibinfo {author} {\bibfnamefont {J.}~\bibnamefont {Dai}}, \bibinfo
  {author} {\bibfnamefont {R.}~\bibnamefont {Wang}}, \bibinfo {author}
  {\bibfnamefont {C.}~\bibnamefont {Wang}},\ and\ \bibinfo {author}
  {\bibfnamefont {W.}~\bibnamefont {Ji}},\ }\bibfield  {title} {\bibinfo
  {title} {{Mechanically and electrically switchable triferroic altermagnet in
  a pentagonal $\mathrm{Fe}{\mathrm{O}}_{2}$ monolayer}},\ }\href
  {https://doi.org/10.1103/ftmr-bh9k} {\bibfield  {journal} {\bibinfo
  {journal} {Phys. Rev. B}\ }\textbf {\bibinfo {volume} {112}},\ \bibinfo
  {pages} {195410} (\bibinfo {year} {2025})}\BibitemShut {NoStop}%
\bibitem [{\citenamefont {Urru}\ \emph {et~al.}(2025)\citenamefont {Urru},
  \citenamefont {Seleznev}, \citenamefont {Teng}, \citenamefont {Park},
  \citenamefont {Reyes-Lillo},\ and\ \citenamefont {Rabe}}]{Urru2025gtype}%
  \BibitemOpen
  \bibfield  {author} {\bibinfo {author} {\bibfnamefont {A.}~\bibnamefont
  {Urru}}, \bibinfo {author} {\bibfnamefont {D.}~\bibnamefont {Seleznev}},
  \bibinfo {author} {\bibfnamefont {Y.}~\bibnamefont {Teng}}, \bibinfo {author}
  {\bibfnamefont {S.~Y.}\ \bibnamefont {Park}}, \bibinfo {author}
  {\bibfnamefont {S.~E.}\ \bibnamefont {Reyes-Lillo}},\ and\ \bibinfo {author}
  {\bibfnamefont {K.~M.}\ \bibnamefont {Rabe}},\ }\bibfield  {title} {\bibinfo
  {title} {{$G$-type antiferromagnetic ${\mathrm{BiFeO}}_{3}$ is a multiferroic
  $g$-wave altermagnet}},\ }\href {https://doi.org/10.1103/v3fg-6smc}
  {\bibfield  {journal} {\bibinfo  {journal} {Phys. Rev. B}\ }\textbf {\bibinfo
  {volume} {112}},\ \bibinfo {pages} {104411} (\bibinfo {year}
  {2025})}\BibitemShut {NoStop}%
\end{thebibliography}%
%

\end{document}